\documentclass[fleqn,usenatbib]{mnras}

\usepackage{newtxtext,newtxmath}
\usepackage[T1]{fontenc}

\DeclareRobustCommand{\VAN}[3]{#2}
\let\VANthebibliography\thebibliography
\def\thebibliography{\DeclareRobustCommand{\VAN}[3]{##3}\VANthebibliography}

\usepackage{graphicx}	% Including figure files
\usepackage{afterpage}   % 図を次のページに配置するためのパッケージ

\usepackage{amsmath}	% Advanced maths commands
\usepackage{CJKutf8}%! To show the Japanese language.

\newcommand{\msun}{{\rm M}_\odot}

\newcommand{\cc}{{\rm cm^{-3}}}
\newcommand{\kms}{{\rm km\,s^{-1}}}
\newcommand{\nh}{n_{\rm H}}
\newcommand{\nth}{n_{\rm th}}
\newcommand{\tth}{t_{\rm th}}
\newcommand{\vrad}{v_{\rm rad}}
\newcommand{\vsv}{v_{\rm SV}}
\newcommand{\vsvsigma}{v_{\rm SV}/\sigma_{\rm SV}}

\newcommand{\Rvir}{R_{\rm v}}
\newcommand{\Mvir}{M_{\rm v}}
\newcommand{\Tvir}{T_{\rm v}}
\newcommand{\fbaryon}{f_{\rm b}}
\newcommand{\Ncore}{N_{\rm c}}
\newcommand{\Mcore}{M_{\rm c}}
\newcommand{\Mcoretot}{M_{\rm c,tot}}
\newcommand{\Mcorefirst}{M_{\rm c,1}}
\newcommand{\Mcoresecond}{M_{\rm c,2}}
\newcommand{\qcore}{q_{\rm c}}
\newcommand{\epsIII}{\epsilon_{\rm III}}
\newcommand{\Mcrit}{M_{\rm crit}}
\newcommand{\Mcool}{M_{\rm cool}}
\newcommand{\vcirc}{v_{\rm circ}}
\graphicspath{{./}{figures/}}

\defcitealias{Hirano2018}{Hirano et al. 2018}
\defcitealias{Hirano2023PaperI}{Paper~I}

\title[First star clusters - II]{Formation of first star clusters under the supersonic gas flow -- II.\\ 
Critical halo mass and core mass function}

\author[S. Hirano et al.]{
Shingo Hirano$^{1,2}$\thanks{E-mail: shingo-hirano@kanagawa-u.ac.jp}\\
$^{1}$Department of Applied Physics, Faculty of Engineering, Kanagawa University, Kanagawa 221-0802, Japan\\
$^{2}$Department of Astronomy, School of Science, University of Tokyo, Tokyo 113-0033, Japan
}

\date{Accepted 2025 May 8. Received 2025 Apr 1; in original form 2025 January 05}
\pubyear{2025}

\begin{document}
\label{firstpage}
\pagerange{\pageref{firstpage}--\pageref{lastpage}}
\maketitle

\begin{CJK}{UTF8}{min}%! To show the Japanese language.

\begin{abstract}
The formation and mass distribution of the first stars depend on various environmental factors in the early universe.
We compare 120 cosmological hydrodynamical simulations to explore how the baryonic streaming velocity (SV) relative to dark matter affects the formation of the first stars.
We vary SV from zero to three times its cosmic root-mean-square value, $\vsvsigma=0-3$, and identify 20 representative halos from cosmological simulations.
For each model, we follow the evolution of a primordial star-forming cloud from the first appearance of a dense core (with gas density > $10^{6}\,\cc$) until 2\,Myr later.
In each model, higher SV systematically delays the formation of primordial clouds, formed inside more massive halos ($10^{5}-10^{7}\,\msun$), and promotes cloud-scale fragmentation and multiple-core formation.
The number and total mass of dense cores increase with increasing SV.
More than half of models with $\vsvsigma \ge 1.5$ form three or more dense cores in a single halo.
In extreme cases, up to 25 cores form at once, which leaves a massive first star cluster.
On the other hand, models with $\vsvsigma \leq 1$ form only one or two cores in a halo.
In addition, HD-cooling is often enabled in models with low SV, especially in low-z, where HD-cooling is enabled in more than 50\% of models.
This leads to the formation of the low-mass first star.
SV shapes the resulting initial mass function of the first stars and plays a critical role in setting the star-forming environment of the first galaxies.
\end{abstract}


%%%%%%%%%%%%%%%%%%%%%%%%%%%%%%%%%%%%%%%%%%%%%%%%%%
%%%%%%%%%%%%%%%%% BODY OF PAPER %%%%%%%%%%%%%%%%%%
%%%%%%%%%%%%%%%%%%%%%%%%%%%%%%%%%%%%%%%%%%%%%%%%%%

%%%%%%%%%%%%%%%%%%%%%%%%%%%%%%%%%%%%%%%%%%%%%%%%%%
\section{Introduction}
\label{sec:intro}
%%%%%%%%%%%%%%%%%%%%%%%%%%%%%%%%%%%%%%%%%%%%%%%%%%

The formation of the first stars, or Population III (Pop III) stars, marks a key transition in cosmic history.
It shapes the earliest stages of galaxy evolution and the chemical enrichment of the universe \citep[see][for a review]{KlessenGlover2023}.
Recent advances in observational facilities, notably the James Webb Space Telescope (JWST) and the Atacama Large Millimeter/submillimeter Array (ALMA), have begun to probe the cosmic dawn, enabling the study of galaxies at redshifts as high as $z=10-15$ \citep[e.g.,][]{Robertson2024, Harikane2024}.
As we enter this ``deep universe era,'' ongoing and planned facilities -- such as the Thirty Meter Telescope (TMT) and the Giant Magellan Telescope (GMT) -- promise even deeper insights into the birth of the first galaxies and their stellar populations.

\begin{figure}
\begin{center}
\includegraphics[width=1.0\linewidth]{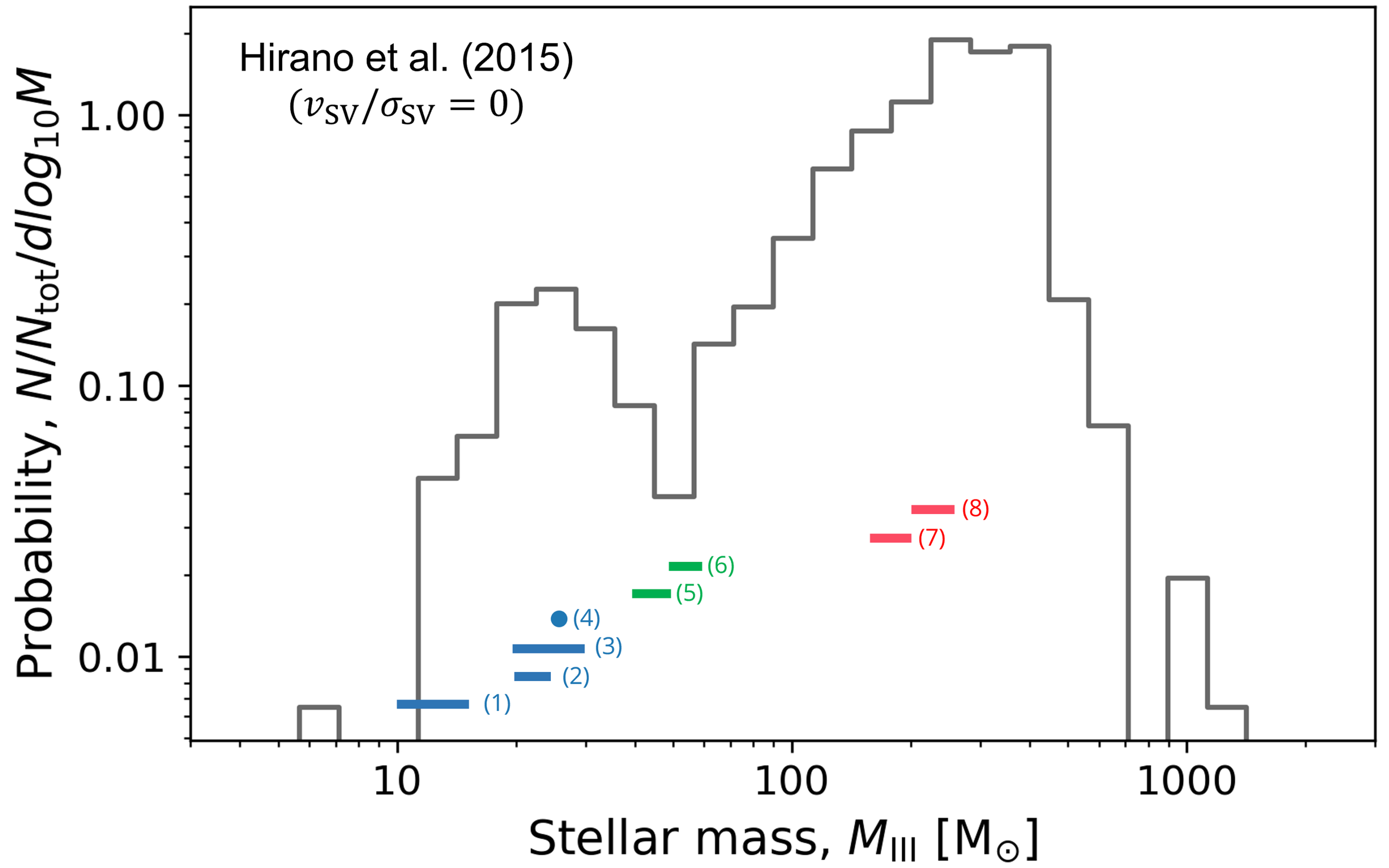}
\end{center}
\caption{
Comparison of the theoretical estimation of the stellar mass distribution of the first stars \citep{Hirano2015} with indirect observational clues regarding the mass range of the parent first star; 
(A) Core-collapse supernovae (CCSN; $10-40\,\msun$): 
(1) LMC-119 \citep{Chiti2024}, 
(2) HE 0020-1741 \citep{Placco2016}, 
(3) SDSS J102915+172927 \citep{Caffau2011, Schneider2012}, 
LAMOST J221750.59+210437.2 \citep{Aoki2018}, 
2MASS J20500194-6613298 \citep{Mardini2024},
(4) SMSS J031300-670839.3 \citep{Keller2014, Bessell2015}, 
(B) Hypernova ($40-60\,\msun$): 
(5) SPLUS J210428.01-004934.2 \citep{Placco2021},
(6) AS0039 \citep{Skuladottir2021},
(C) Pair-instability supernovae (PISN; $140-260\,\msun$): 
(7) LAMOST J1010+2358 \citep{Xing2023},
(8) SDSS J001820.5-093939.2 \citep{Aoki2014}.
}
\label{fig:EMP}
\end{figure}

Despite these observational leaps, the direct detection of the first stars remains elusive.
Instead, their initial mass function (IMF) constraints rely on indirect evidence.
Extremely metal-poor (EMP) stars in the Milky Way and dwarf galaxies carry the chemical imprint of the first supernovae, allowing researchers to infer the IMF of the earliest stellar generations \citep[e.g.,][and also see Figure~\ref{fig:EMP}]{Keller2014, Bessell2015, Rossi2024}.
However, these data remain incomplete, and certain mass ranges, such as those leading to direct black hole formation, are not directly constrained.
Thus, theoretical models and numerical simulations are central to understanding how the first stars formed and how their mass distribution emerged.

\begin{figure*}
\begin{center}
\includegraphics[width=0.8\linewidth]{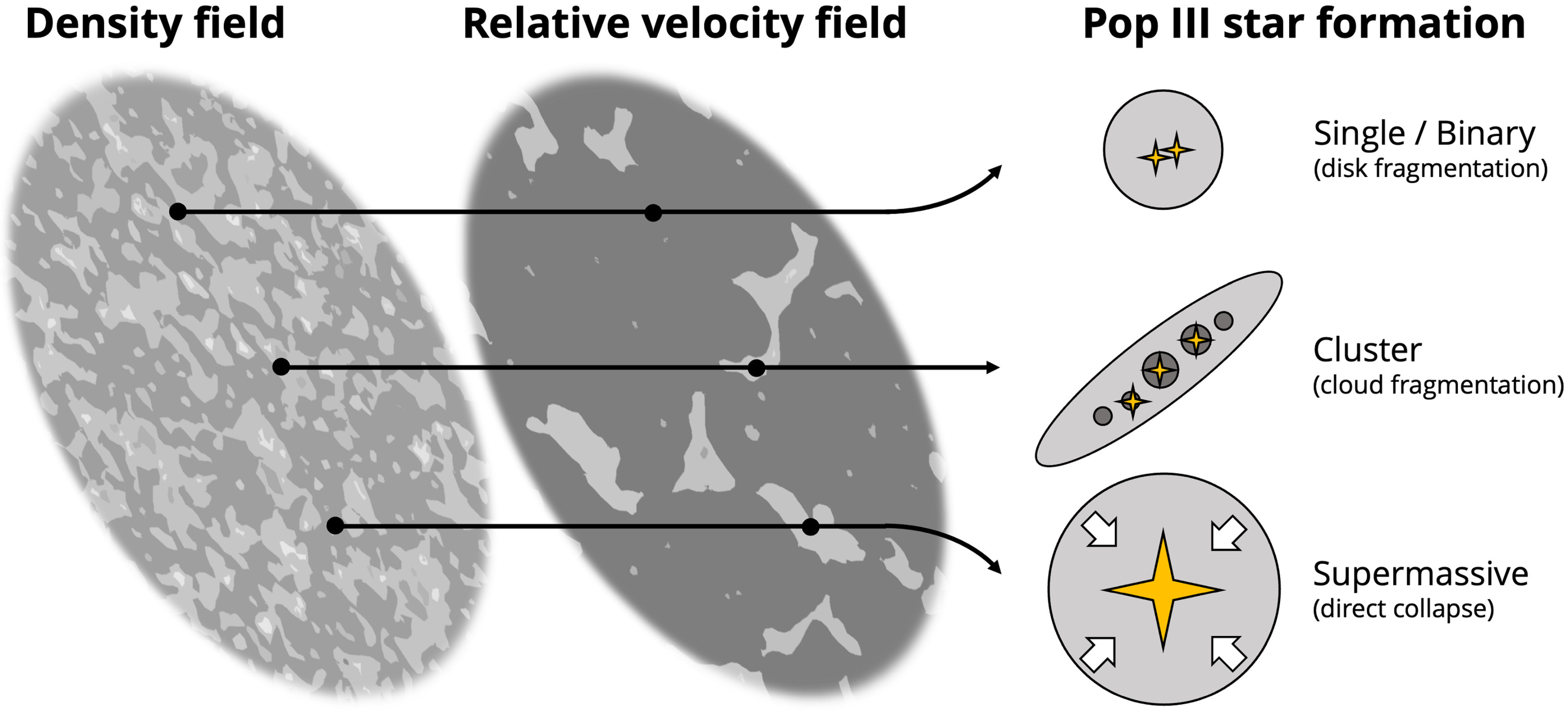}
\end{center}
\caption{
Schematic diagram of the dependence of the first star formation process on baryonic supersonic motions relative to dark matter in the early universe.
While the typical first star formation process proceeds in regions with negligible relative velocities, multiple star-forming gas clouds can form due to cloud-scale fragmentation in areas with low relative velocities, and supermassive first stars can form from the direct collapse process in regions with high relative velocities.
}
\label{fig:SVeffect}
\end{figure*}

One of the external factors influencing the primordial star formation process is the streaming velocity (SV): the relative, supersonic motion between baryonic gas and dark matter (DM) imprinted at cosmic recombination \citep[][]{Tseliakhovich2010, Fialkov2014}.
This effect is not merely a subtle perturbation; it can qualitatively alter the conditions of star formation.
Enhanced SV values delay the onset of gas collapse, shift star formation to more massive DM halos, increase the characteristic gas mass scale, and thereby shape the IMF of the first stars \citep[e.g.,][]{Greif2011sv, Stacy2011, Naoz2014}.
Figure~\ref{fig:SVeffect} visualizes the impact of SV on the first star formation based on a set of cosmological simulations \citep[]{Hirano2017smbh, Hirano2018}.
Moreover, SV can give rise to unique phenomena, such as Supersonically Induced Gas Objects (SIGOs), which form in regions where SV suppresses DM clustering and gas assembles into dense, star-forming structures with low DM content \citep[e.g.,][]{Chiou2018, Chiou2019}.
SIGOs may represent a new pathway for globular cluster formation, bridging conditions in the high-redshift universe to the present-day population of globular clusters.

While numerous simulations have explored the baseline scenario of the first star formation under standard conditions without significant SV, recent studies have begun to systematically incorporate SV and other environmental parameters, such as external radiation \citep{Schauer2021SV+UV, Kulkarni2021}.
Among these works, we focus on how SV influences the formation of primordial star-forming gas clouds within DM halos.
Our prior work \citep[][hereafter \citetalias{Hirano2023PaperI}]{Hirano2023PaperI} introduced a methodology for characterizing the formation of the first star clusters under different SV.

In this second paper, we systematically investigate the SV effect on the first star formation.
Extending our previous study \citep{Hirano2023PaperI}, we perform a parameter survey by selecting 20 representative DM halos and applying six different SV amplitudes (0, 1, 1.5, 2, 2.5, and 3 times the cosmic root-mean-square value).
By following the evolution of primordial gas clouds until 2\,Myr after the cloud collapse, we analyze the formation of multiple dense cores where $n \geq 10^6\,\cc$ within a single halo, derive the core mass function, and identify the critical halo mass scales that govern when and where first star clusters appear.
This results in 120 distinct zoom-in simulations that allow us to explore how SV influences the onset of star formation and the mass distribution of primordial gas clouds.

This paper is organized as follows.
Section~\ref{sec:method} describes simulation methods and initial conditions.
Section~\ref{sec:res} presents the results of our parameter survey, focusing on the critical halo mass and the resulting core mass functions.
Section~\ref{sec:dis} discusses the implications of our findings for the IMF of Pop~III stars and the formation of first star clusters.
Finally, Section~\ref{sec:con} summarizes our conclusions and outlines directions for future work in this series.

%%%%%%%%%%%%%%%%%%%%%%%%%%%%%%%%%%%%%%%%%%%%%%%%%%
%%%%%%%%%%%%%%%%%%%%%%%%%%%%%%%%%%%%%%%%%%%%%%%%%%

%%%%%%%%%%%%%%%%%%%%%%%%%%%%%%%%%%%%%%%%%%%%%%%%%%
%%%%%%%%%%%%%%%%%%%%%%%%%%%%%%%%%%%%%%%%%%%%%%%%%%

%%%%%%%%%%%%%%%%%%%%%%%%%%%%%%%%%%%%%%%%%%%%%%%%%%
\section{Numerical methodology}
\label{sec:method}
%%%%%%%%%%%%%%%%%%%%%%%%%%%%%%%%%%%%%%%%%%%%%%%%%%

We perform a set of three-dimensional cosmological hydrodynamical simulations to study the dependence of first star formation under the baryonic streaming motions in the early universe.
The simulation setup follows \citetalias{Hirano2023PaperI}. However, we reduce the numerical resolution to simulate the long-term evolution of the primordial star-forming gas clouds.
Furthermore, we increase the number of host haloes for which we study the effects of the baryonic streaming motion from seven to twenty samples to be able to discuss the dependence on the formation environment in more detail.

%%%%%%%%%%%%%%%%%%%%%%%%%%%%%%%%%%%%%%%%%%%%%%%%%%
\subsection{Initial condition}
\label{sec:method:ics}
%%%%%%%%%%%%%%%%%%%%%%%%%%%%%%%%%%%%%%%%%%%%%%%%%%

We first select 20 DM halos with masses of $10^{5}-10^{6}\,\msun$ from large-scale cosmological simulations that assume no initial streaming velocity.
We then apply a hierarchical zoom-in technique to refine each halo step by step, ultimately achieving a mass resolution of about $0.03\,\msun$ for gas particles.
Finally, we introduce a uniform relative velocity between DM and baryons ($\vsvsigma=0-3$) for each zoom-in initial condition, resulting in a total of 120 distinct models.

We first use the public code \texttt{MUSIC} \citep{HahnAbel2011} to generate the base cosmological ICs for a comoving volume of $L_{\rm box} = 10\,h^{-1}$\,comoving megaparsec (cMpc) per side at $z_{\rm ini}=499$.
Our adopted $\Lambda$CDM cosmology \citep{PLANCK2018} has $\Omega_{\rm m}=0.31$, $\Omega_{\rm b}=0.048$, $\Omega_\Lambda=0.69$, $H_0=68\,\kms\,{\rm Mpc}^{-1}$, $\sigma_8=0.83$, and $n_{\rm s}=0.96$.

We then run cosmological $N$-body/hydrodynamics simulations using \texttt{GADGET-3} \citep[][]{Springel2005}.
We identify the first DM halo that forms in each cosmological IC without streaming motion and re-simulate it at higher resolution using hierarchical zoom-in ICs generated by \texttt{MUSIC}.
With five levels of refinement, the effective resolution improves from $512^3$ to $16384^3$, reducing the DM particle mass from $9.426\times10^5\,\msun$ to $28.76\,\msun$.
We obtain 20 such zoomed-in halos.
Seven of these correspond to Halos A-G in \citetalias{Hirano2023PaperI} (see the caption of Table~\ref{table:each}).

We introduce a uniform initial relative velocity between the DM and baryonic components along the $x$-axis to model the baryonic streaming motion.
Because the coherence length of the streaming velocity field extends over a few megaparsecs, well beyond the scale of our DM halos, assuming a uniform velocity is appropriate.
Under the baryons-trace-dark-matter (BTD) approximation \citep{Park2020btd, Park2021btd}, we assume that the initial baryon density matches the DM density distribution.
We generate six sets of ICs, each sharing the same density phase but differing in their initial streaming velocity: $\vsvsigma =0$, 1, 1.5, 2, 2.5, and 3, normalized by the root-mean-square velocity $\sigma_{\rm SV}(z) = \sigma_{\rm SV}^{\rm rec}(1+z)/(1+z_{\rm rec}) = 13.76\,\kms$ at $z_{\rm ini} = 499$.
This value is derived from $\sigma_{\rm SV}^{\rm rec}=30\,\kms$ at the cosmic recombination era ($z_{\rm rec}=1089$).

In total, we have 120 models: twenty DM halos combined with six different streaming velocities.
Table~\ref{table:each} lists these models, where the model names are defined by a halo ID (I01-I20) and a velocity label (V00, V10, V15, V20, V25, and V30).

%%%%%%%%%%%%%%%%%%%%%%%%%%%%%%%%%%%%%%%%%%%%%%%%%%
\begin{figure*}
\begin{center}
\includegraphics[width=1.0\linewidth]{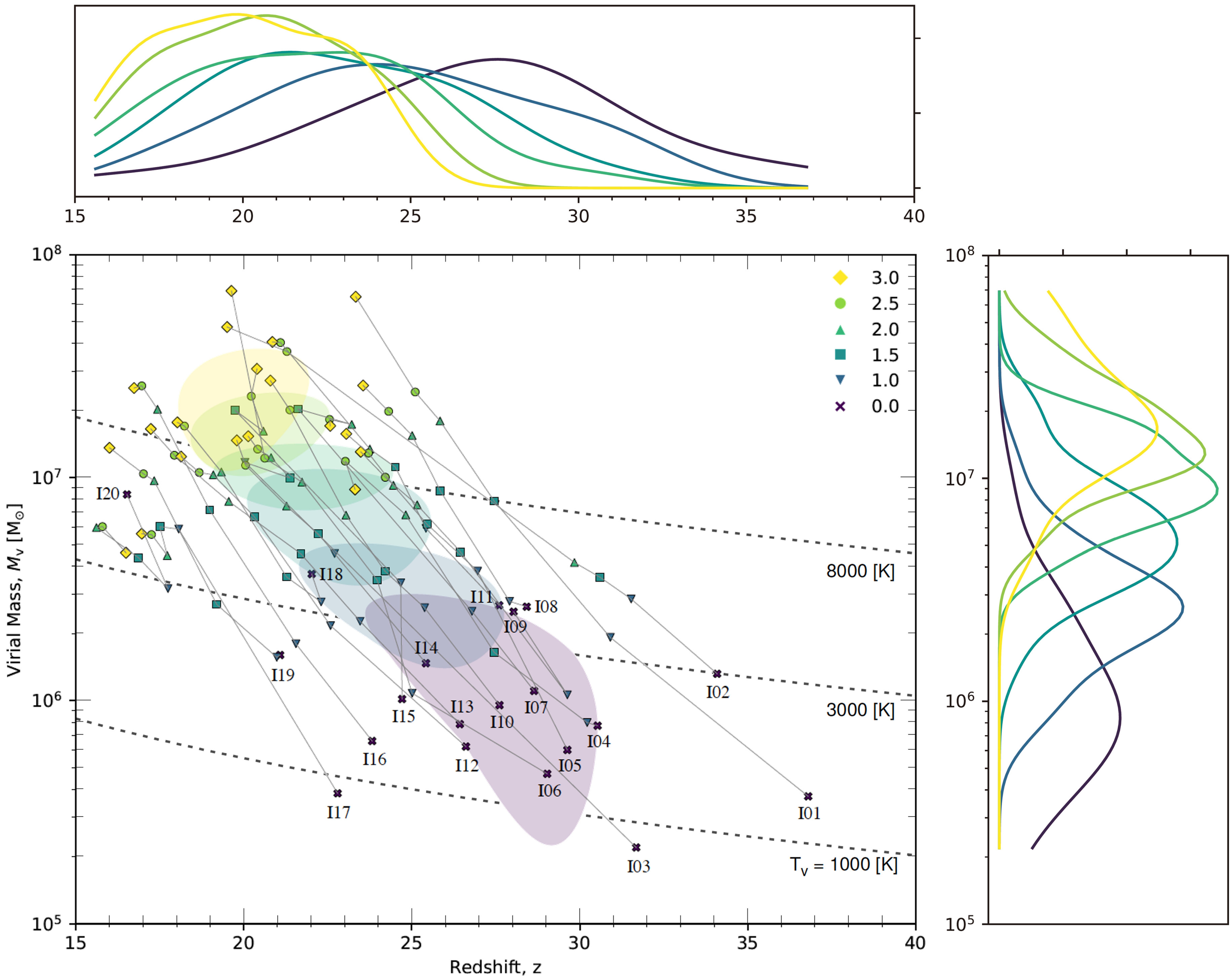}
\end{center}
\caption{
Distribution of redshift and virial halo mass ($z$-$\Mvir$ diagram) for $120$ models.
The colours and symbols correspond to the magnitude of the initial streaming velocity (V00-V30 with $\vsvsigma=0$, 1, 1.5, 2, 2.5, and 3).
The solid lines connect models that examined identical density fluctuations (I01-I20).
We compute the mean and variance of the data for each streaming velocity and generate histograms using the Kernel Density Estimation (top and right panels).
The coloured areas in the central panel show $z$-$\Mvir$ distributions for each initial streaming velocity.
The dashed lines show the virial masses for three different virial temperatures, $\Tvir = 1000$, $3000$, and $8000$\,K (Equation~\ref{eq:Mvir}).
}
\label{fig:z-M_each}
\end{figure*}
%%%%%%%%%%%%%%%%%%%%%%%%%%%%%%%%%%%%%%%%%%%%%%%%%%

%%%%%%%%%%%%%%%%%%%%%%%%%%%%%%%%%%%%%%%%%%%%%%%%%%
\begin{table*}
\centering
\caption{Simulation results averaged for each classified model}
\label{table:average}
\begin{tabular}{@{}lcccccllrcrrc@{}}
\hline
Class & $\vsvsigma$ & $\overline{z}$ & $\overline{\Rvir}$ & $\overline{\Mvir}$ & $\overline{\fbaryon}$ & $N_{\rm HD}/N$& $\overline{\Ncore}$ & $\overline{\Mcoretot}$ & $\overline{\epsIII}$ & $\overline{\Mcorefirst}$ & $\overline{\Mcoresecond}$ & $\overline{\qcore}$ \\
& & & (pc) & ($\msun$) & & & & ($\msun$) & & ($\msun$) & ($\msun$) & \\
\hline
\multicolumn{12}{l}{({\it All})} \\
A00 & 0.0 & 27.07 & 122.3 & $1.072\times10^6$ & 0.133 & 0.20 & 1.55 &  6445 & 0.0453 &  5856 &  975 & 0.166 \\
A10 & 1.0 & 24.69 & 178.9 & $2.637\times10^6$ & 0.129 & 0.35 & 1.60 &  8729 & 0.0257 &  7525 & 4072 & 0.541 \\
A15 & 1.5 & 22.83 & 251.2 & $5.838\times10^6$ & 0.126 & 0.25 & 3.25 & 13794 & 0.0188 & 10638 & 2597 & 0.244 \\
A20 & 2.0 & 21.78 & 314.4 & $9.695\times10^6$ & 0.123 & 0.10 & 5.20 & 19940 & 0.0167 & 14331 & 4221 & 0.295 \\
A25 & 2.5 & 20.49 & 380.2 & $1.503\times10^7$ & 0.122 & 0    & 4.60 & 22876 & 0.0125 & 16284 & 4701 & 0.289 \\
A30 & 3.0 & 20.00 & 421.7 & $1.908\times10^7$ & 0.121 & 0.05 & 4.65 & 20514 & 0.0089 & 14284 & 3567 & 0.250 \\
\multicolumn{12}{l}{({\it High})} \\
H00 & 0.0 & 30.99 & 120.2 & $1.540\times10^6$ & 0.124 & 0    & 1    & 11407 & 0.0598 & 11407 &    - &     - \\
H10 & 1.0 & 29.08 & 147.9 & $2.491\times10^6$ & 0.121 & 0.20 & 1.20 & 11043 & 0.0367 & 10149 & 4089 & 0.403 \\
H15 & 1.5 & 26.97 & 218.8 & $6.610\times10^6$ & 0.114 & 0.20 & 4.60 & 19929 & 0.0266 & 14374 & 1861 & 0.129 \\
H20 & 2.0 & 25.92 & 269.2 & $1.031\times10^7$ & 0.113 & 0    & 7.60 & 30962 & 0.0267 & 18326 & 5262 & 0.287 \\
H25 & 2.5 & 23.20 & 389.0 & $2.345\times10^7$ & 0.119 & 0    & 5.60 & 33337 & 0.0120 & 17958 & 6815 & 0.380 \\
H30 & 3.0 & 22.14 & 457.1 & $3.345\times10^7$ & 0.125 & 0    & 7.60 & 33322 & 0.0080 & 20148 & 7594 & 0.377 \\
\multicolumn{12}{l}{({\it Middle})} \\
M00 & 0.0 & 28.41 & 101.4 & $7.240\times10^5$ & 0.128 & 0.25 & 1.88 &  4198 & 0.0453 &  3670 &  668 & 0.182 \\
M10 & 1.0 & 25.22 & 177.8 & $2.808\times10^6$ & 0.123 & 0.25 & 2.00 & 10694 & 0.0308 &  8468 & 6448 & 0.761 \\
M15 & 1.5 & 23.25 & 254.8 & $6.458\times10^6$ & 0.122 & 0.38 & 4.25 & 15663 & 0.0199 & 10604 & 2937 & 0.277 \\
M20 & 2.0 & 22.49 & 307.3 & $1.005\times10^7$ & 0.120 & 0.13 & 4.88 & 20360 & 0.0170 & 16171 & 4291 & 0.265 \\
M25 & 2.5 & 21.49 & 360.0 & $1.488\times10^7$ & 0.119 & 0    & 6.00 & 24086 & 0.0136 & 16265 & 7174 & 0.441 \\
M30 & 3.0 & 21.21 & 409.7 & $2.045\times10^7$ & 0.114 & 0.13 & 5.00 & 30311 & 0.0130 & 21134 & 4765 & 0.225 \\
\multicolumn{12}{l}{({\it Low})} \\
L00 & 0.0 & 22.75 & 153.4 & $1.296\times10^6$ & 0.144 & 0.29 & 1.57 &  6998 & 0.0374 &  6204 & 2508 & 0.404 \\
L10 & 1.0 & 20.95 & 206.2 & $2.557\times10^6$ & 0.140 & 0.57 & 1.43 &  5852 & 0.0163 &  5310 & 1621 & 0.305 \\
L15 & 1.5 & 19.40 & 272.7 & $4.761\times10^6$ & 0.139 & 0.14 & 1.14 &  9172 & 0.0139 &  8613 & 6579 & 0.764 \\
L20 & 2.0 & 18.01 & 360.7 & $8.905\times10^6$ & 0.136 & 0.14 & 3.86 & 14220 & 0.0118 & 10472 & 3332 & 0.318 \\
L25 & 2.5 & 17.41 & 398.1 & $1.106\times10^7$ & 0.127 & 0    & 2.29 & 16480 & 0.0118 & 15205 & 2265 & 0.149 \\
L30 & 3.0 & 17.08 & 411.4 & $1.181\times10^7$ & 0.125 & 0    & 2.14 &  9285 & 0.0063 &  7140 & 1085 & 0.152 \\
\multicolumn{12}{l}{({\it HD})} \\
D00 & 0.0 & 25.99 & 122.3 & $1.013\times10^6$ & 0.140 & -    & 2.75 &  4406 & 0.0310 &  3555 &  581 & 0.164 \\
D10 & 1.0 & 23.78 & 172.1 & $2.181\times10^6$ & 0.128 & -    & 1    &  3947 & 0.0142 &  3947 &    - &     - \\
D15 & 1.5 & 24.30 & 213.8 & $4.373\times10^6$ & 0.125 & -    & 2.80 &  6384 & 0.0117 &  5382 &  597 & 0.111 \\
D20 & 2.0 & 21.19 & 298.5 & $8.470\times10^6$ & 0.136 & -    & 6.50 & 11563 & 0.0101 &  8884 &  706 & 0.080 \\
D25 & 2.5 &     - &     - &                 - &     - & -    &    - &     - &      - &     - &    - &     - \\
D30 & 3.0 & 22.57 & 354.8 & $1.705\times10^7$ & 0.109 & -    & 1    &  7494 & 0.0040 &  7494 &    - &     - \\
\hline
\end{tabular}
\begin{flushleft}
{\it Notes.} Column 1: classification name.
Column 2: relative streaming velocity normalized by the root-mean-square value ($\vsvsigma$).
Column 3: redshift ($z$) when the gas number density firstly reaches $\nh = 10^6\,\cc$.
Columns 4-6: radius ($\Rvir$), mass ($\Mvir$), and baryon fraction ($\fbaryon$) at the virial scale.
Column 7: proportion of the HD-cooling models that meet the abundance ratio criterion $f_{\rm HD}/f_{\rm H_2} \geq 10^{-3}$ at the end of the calculation $\tth=2$\,Myr.
Column 8: number of cores ($\Ncore$).
Column 9: the total mass of cores ($\Mcoretot$).
Column 10: mass conversion efficiency ($\epsIII=\Mcoretot/(\fbaryon \Mvir)$).
Columns 11 and 12: mass of the primary and secondary core ($\Mcorefirst$ and $\Mcoresecond$).
Column 13: mass ratio of the primary and secondary cores ($\qcore=\Mcoresecond/\Mcorefirst$).
Table~\ref{table:each} shows all data for each model.
We average the results of all models ({\it All}), three classified groups ({\it High}, {\it Middle}, and {\it Low}), and models with HD-cooling enabled ({\it HD}) for $6$ different initial streaming velocities: ({\it High}) I01, I02, I08, I09, I11, ({\it Middle}) I03-07, I10, I14, I15, ({\it Low}) I12, I13, I16-20, and ({\it HD}) see column~7 in Table~\ref{table:each}.
There is no data in columns 12 and 13 for H00, D10, and D30 because none of the models belonging to them have a secondary core.
\end{flushleft}
\end{table*}
%%%%%%%%%%%%%%%%%%%%%%%%%%%%%%%%%%%%%%%%%%%%%%%%%%

%%%%%%%%%%%%%%%%%%%%%%%%%%%%%%%%%%%%%%%%%%%%%%%%%%
\subsection{Cosmological simulation}
\label{sec:method:sim}
%%%%%%%%%%%%%%%%%%%%%%%%%%%%%%%%%%%%%%%%%%%%%%%%%%

We perform the cosmological simulations with a modified version of the parallel $N$-body/smoothed particle hydrodynamics (SPH) code \texttt{GADGET-3} \citep{Springel2005}, adapted for metal-free star formation \citep{Hirano2018}, and including detailed non-equilibrium chemistry of 14 species (e$^-$, H, H$^+$, H$^-$, He, He$^+$, He$^{++}$, H$_2$, H$_2^+$, D, D$^+$, HD, HD$^+$, HD$^-$) as in \citet{Yoshida2007, Yoshida2008}.
To follow gas collapse down to $\nth=10^6\,\cc$, we apply a hierarchical refinement scheme that ensures the local Jeans length is always resolved.
Specifically, we require that 15 times the smoothing length is less than the local Jeans length (or about $1000$ SPH particle mass is less than the local Jeans mass), and we increase resolution through the particle splitting technique \citep{KitsionasWhitworth2002}.
This yields minimum particle masses of $m_{\rm DM,min}=0.1439\,\msun$ for DM and $m_{\rm gas,min}=0.02636\,\msun$ for gas.

We follow the evolution for 2\,Myr after the gas cloud first reaches $\nth=10^6\,\cc$.
We define $\tth=0$\,yr as the time when the collapsing gas cloud in each model reaches this threshold density.
To enable such long-term evolution, we adopt an opaque core approach \citep{HiranoBromm2017}, artificially suppressing the gas cooling rate above $\nth$:
\begin{equation}
  \Lambda_{\rm red} = \beta_{\rm esc,art} \cdot \Lambda_{\rm thin} \, ,
  \label{eq:Lambda}
\end{equation}
with an artificial escape fraction and an artificial optical depth as
\begin{equation}
  \beta_{\rm esc,art} = \frac{1-\exp(\tau_{\rm art})}{\tau_{\rm art}}, \tau_{\rm art} = \left( \frac{n}{\nth} \right)^2 \, .
  \label{eq:EscapeFraction}
\end{equation}
This enhancement in effective optical depth halts further collapse, allowing us to study the large-scale evolution of the star-forming region.
We also omit unnecessary chemistry calculations for gas particles above $n_{\rm th}$.
Note that 2\,Myr is shorter than the typical lifetime of a first star \citep{Schaerer2002}, so no supernova feedback affects our halos during this period.

%%%%%%%%%%%%%%%%%%%%%%%%%%%%%%%%%%%%%%%%%%%%%%%%%%
\section{First star formation under the supersonic gas flow}
\label{sec:res}
%%%%%%%%%%%%%%%%%%%%%%%%%%%%%%%%%%%%%%%%%%%%%%%%%%

Figure~\ref{fig:z-M_each} is the distribution of redshift and virial halo mass ($z$-$\Mvir$ diagram) when the maximum density of the collapsing gas cloud first reaches $\nth = 10^6\,\cc$ (defined as $\tth=0$\,yr).
The sub-panels show probability density distributions of $z$ (top) and $\Mvir$ (right) for each streaming velocity ($\vsv$).
At higher $\vsv$, the gravitational collapse of the primordial star-forming cloud delays (resulting in a decrease in $z$ as shown in the top panel), and the host DM halo grows in mass (leading to an increase in $\Mvir$ as shown in the right panel).
As a result, in the $z$-$\Mvir$ diagram, the models move from the bottom-right to the top-left with increasing $\vsv$, as confirmed by previous studies.
In addition to the SV dependence, there is another orthogonal variation in the $z$–$\Mvir$ distribution that arises from differences in the magnitude of the primordial density fluctuations that produced the halo (I01–I20).
This discrepancy is associated with the dynamic state of the halo, which determines the occurrence of the cloud collapse.
If the accretion rate along the DM lanes is high or if minihalo mergers occur (the ``violent merger delay'' scenario), the kinetic energy of the DM (and the baryons) increases, delaying gas cloud collapse even when the gas temperature exceeds the threshold ($\Tvir \sim 1000-2000$\,K) necessary for H$_2$ formation and cooling.

\begin{figure}
\begin{center}
\includegraphics[width=1.0\linewidth]{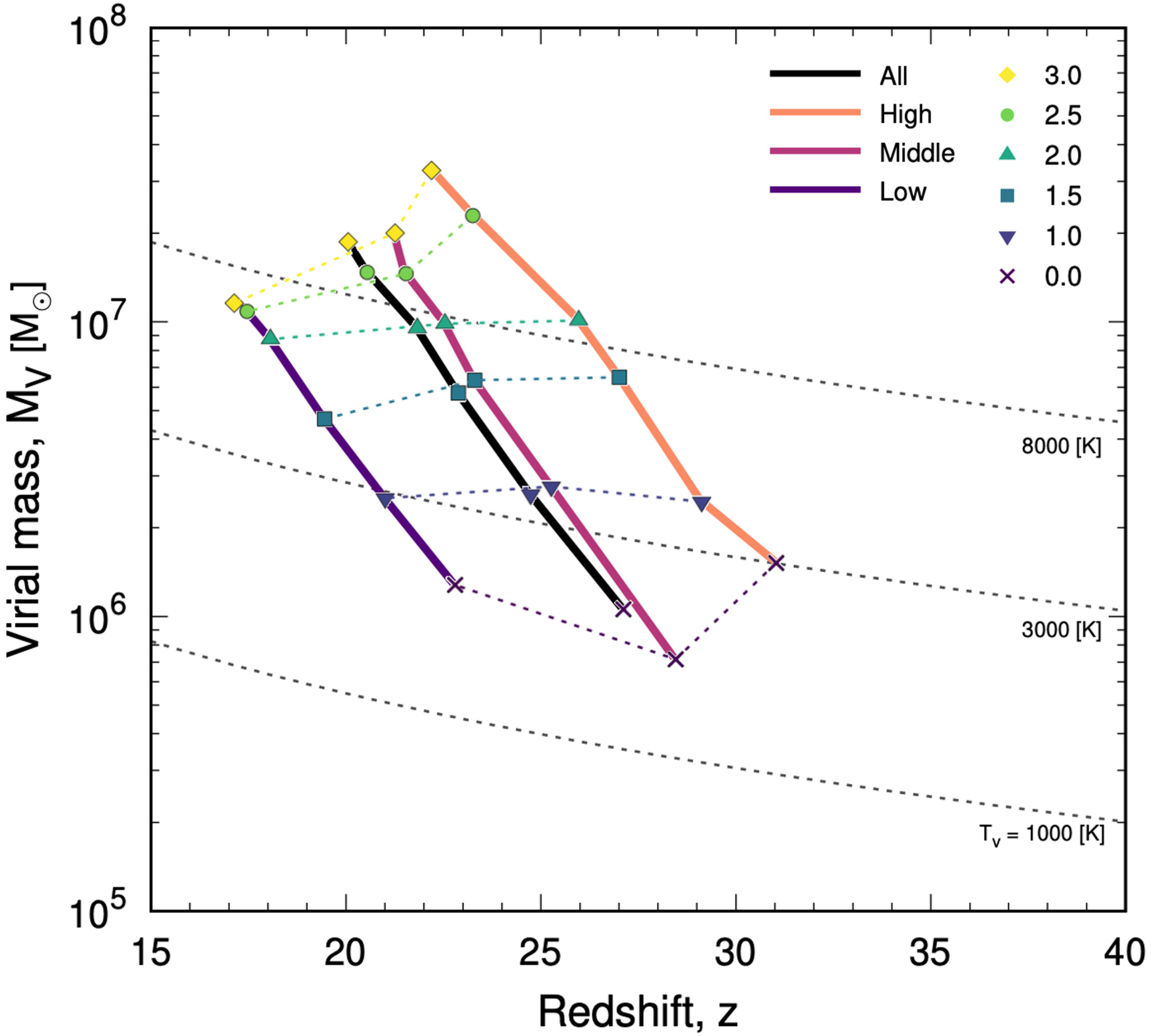}
\end{center}
\caption{
$z$-$\Mvir$ diagram as Figure~\ref{fig:z-M_each} but for averaged results: {\it All} for all models and {\it High}, {\it Middle}, and {\it Low} for the top-right, middle, and bottom-left populations on Figure~\ref{fig:z-M_each} (see also the caption of Table~\ref{table:average}).
}
\label{fig:z-M_average}
\end{figure}

Table~\ref{table:each} summarises the results for 120 models.
To investigate the statistical properties of the SV dependence on the first star formation, we average the analysis results for each SV value across the five classes as Table~\ref{table:average}:
\begin{itemize}
    \item {\it All} shows the averaged values of all models for each SV value.
    \item {\it High}, {\it Middle}, and {\it Low} show averaged values of three groups, the top-right, middle, and bottom-left populations on $z$-$\Mvir$ diagram (Figure~\ref{fig:z-M_each}) to study the dependence of the magnitude of primordial density fluctuation.
    \item {\it HD} shows the averaged values of models in which the hydrogen deuteride (HD)-cooling is effective.
\end{itemize}
Figure~\ref{fig:z-M_average} shows the $z$-$\Mvir$ diagram for {\it All}, {\it High}, {\it Middle}, and {\it Low} classes.

In the following subsections, we show the SV dependence of the physical quantities at three scales: the virial DM halo, which is the gravitationally bound system (Section~\ref{sec:res:virial}), the Jeans gas cloud, which is unstable to the gravitational collapse (Section~\ref{sec:res:jeans}), and the dense core, arbitrarily defined by the maximum numerical resolution of this study, $\nth$, where the first star(s) form (Section~\ref{sec:res:CMF}).

%%%%%%%%%%%%%%%%%%%%%%%%%%%%%%%%%%%%%%%%%%%%%%%%%%
%%%%%%%%%%%%%%%%%%%%%%%%%%%%%%%%%%%%%%%%%%%%%%%%%%

%%%%%%%%%%%%%%%%%%%%%%%%%%%%%%%%%%%%%%%%%%%%%%%%%%
\begin{figure}
\begin{center}
\includegraphics[width=1.0\linewidth]{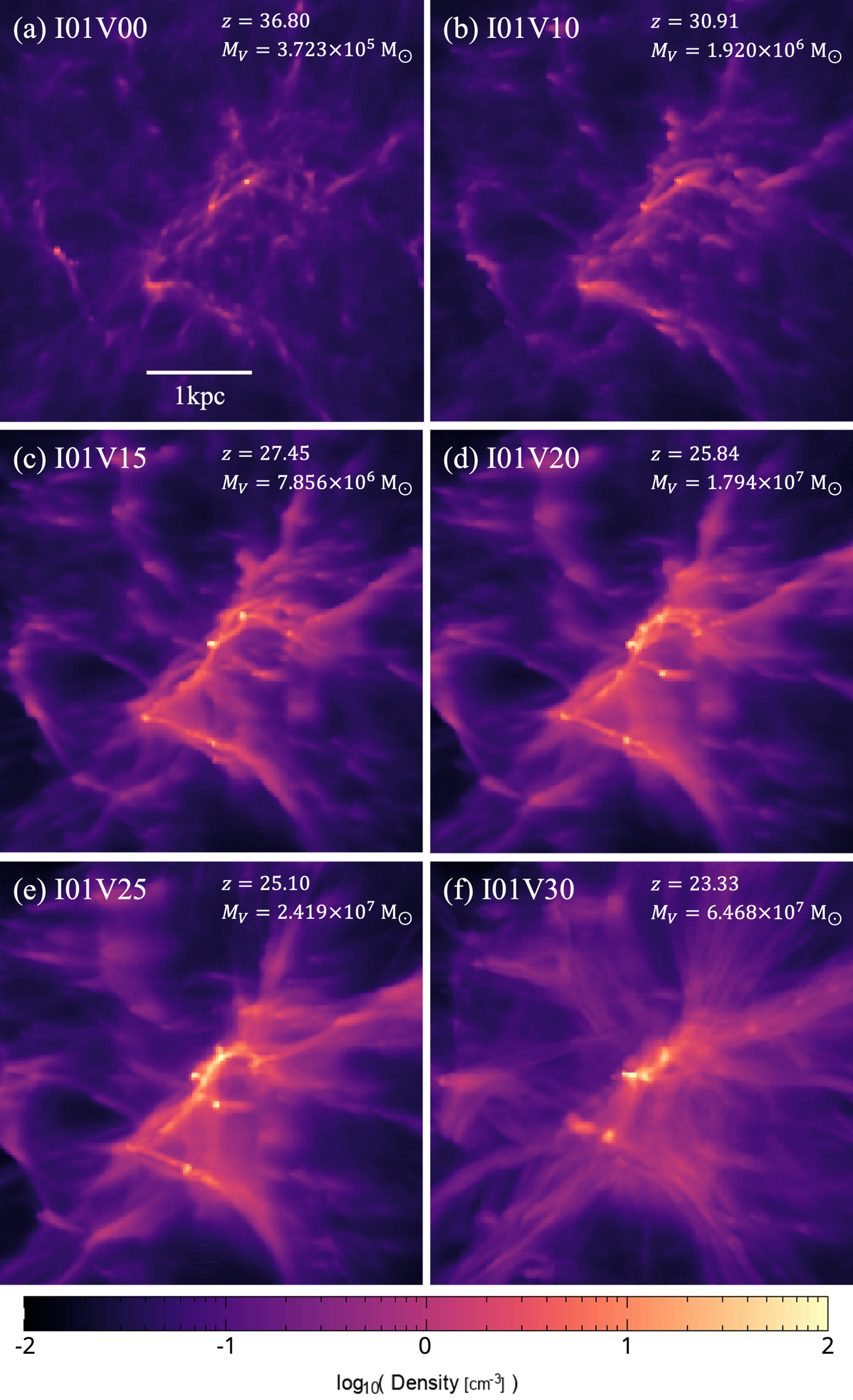}
\end{center}
\caption{
Projected density distribution around the density centre at the completion of gas cloud contraction (when $\tth = 0$\,yr) of I01 models.
The direction of the initial relative velocity between DM and gas components is aligned with the horizontal axis (from left to right) in the figure.
}
\label{fig:2dmap}
\end{figure}
%%%%%%%%%%%%%%%%%%%%%%%%%%%%%%%%%%%%%%%%%%%%%%%%%%

%%%%%%%%%%%%%%%%%%%%%%%%%%%%%%%%%%%%%%%%%%%%%%%%%%
\begin{figure*}
\begin{center}
\includegraphics[width=1.0\linewidth]{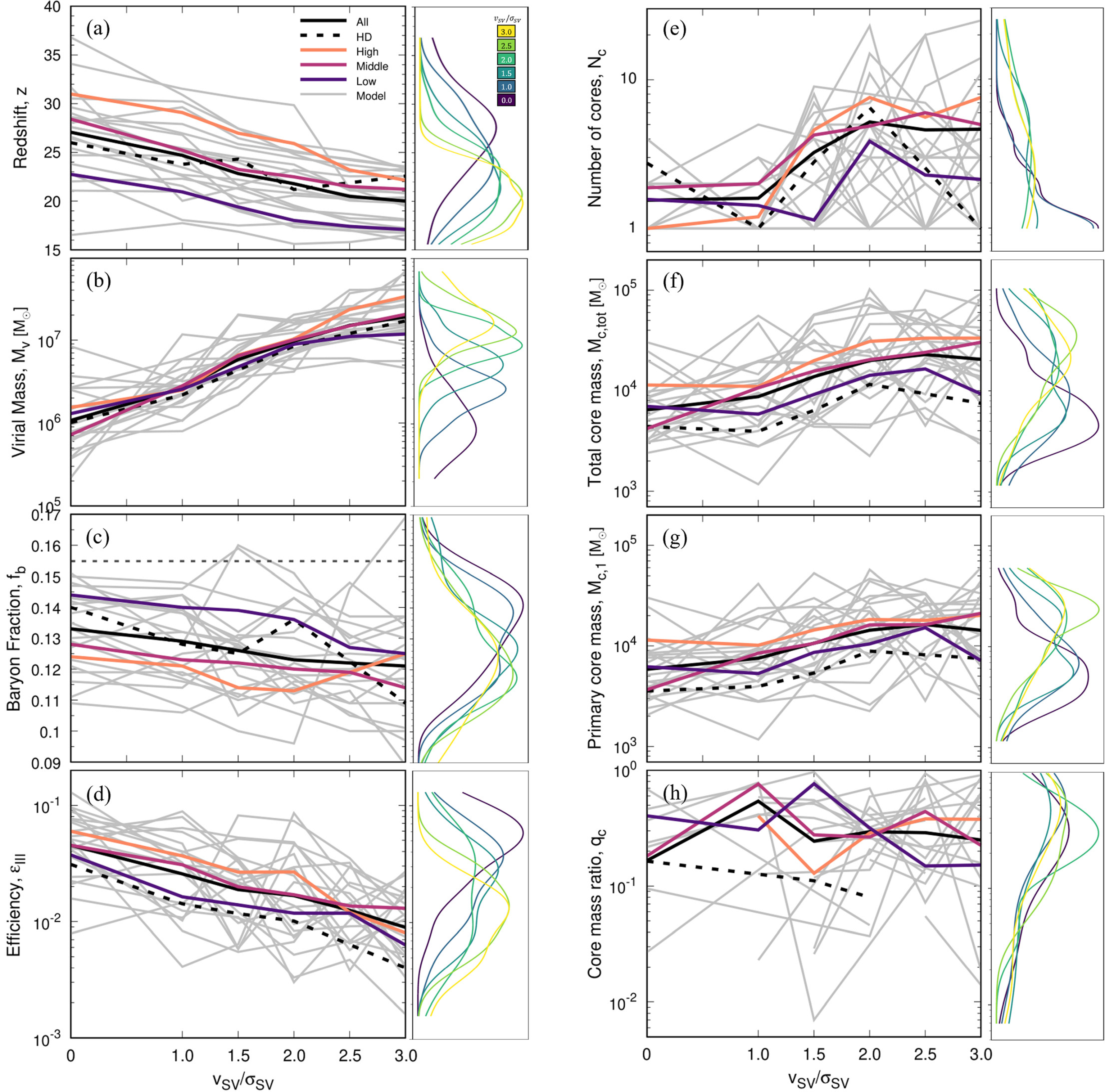}
\end{center}
\caption{
Initial streaming velocity dependence of the calculation results.
Panels: (a) formation redshift, (b) halo mass, (c) baryon fraction (the horizontal dotted line shows the cosmic mean $\Omega_{\rm b}/\Omega_{\rm m}=0.15484$), (d) mass conversion efficiency, (e) number of cores, (f) total mass of cores, (g) primary core mass, and (h) mass ratio of the primary and secondary cores.
The solid and dashed lines show the results averaged over all models ({\it All}) and models with HD-cooling enabled ({\it HD}).
The dashed line in the panel (h) ends up at $\vsvsigma=2.0$ because some parameters have no corresponding model (Table~\ref{table:average}).
The coloured lines show the averaged results for three classified groups: {\it High}, {\it Middle}, and {\it Low}.
The grey lines show the results for each series (I01-I20).
The small panel on the right shows the probability density distribution for each initial streaming velocity as in Figure~\ref{fig:z-M_each}.
}
\label{fig:Vsv-HaloCloud}
\end{figure*}
%%%%%%%%%%%%%%%%%%%%%%%%%%%%%%%%%%%%%%%%%%%%%%%%%%

%%%%%%%%%%%%%%%%%%%%%%%%%%%%%%%%%%%%%%%%%%%%%%%%%%
\subsection{Virial dark matter halo}
\label{sec:res:virial}
%%%%%%%%%%%%%%%%%%%%%%%%%%%%%%%%%%%%%%%%%%%%%%%%%%

We begin by examining how the baryonic SV affects the formation epoch and mass scale of the first star-forming DM halos, which are essential for constructing semi-analytic models and linking initial conditions to subsequent star formation events.
Figure~\ref{fig:z-M_each} is the distribution of the redshift ($z$) and virial halo mass ($\Mvir$) at the onset of gas cooling and collapse for all models in this study.
For models without SV effect ($\vsvsigma=0$; V00), our sample of 20 models spans broad ranges in formation times ($z=16.52-36.80$) and halo masses ($\Mvir=2.195\times10^5 - 8.400\times10^6\,\msun$).
These halos have the virial temperatures $\Tvir \sim 1000 - 3000$\,K, consistent with standard H$_{2}$-cooling minihalos,
\begin{eqnarray}
    \Mvir = 8.120\times10^5\,\msun h^{-1} \left( \frac{\Tvir}{2000\,{\rm K}} \right)^{3/2} \left( \frac{1+z}{25} \right)^{-3/2} \ .
    \label{eq:Mvir}
\end{eqnarray}
On average, the formation epoch is $\overline{z}=27.07$ and the virial mass $\overline{\Mvir}=1.072\times10^6\,\msun$ for the no-SV models (Class A00 in Table~\ref{table:average}).

Introducing SV affects the conditions under which gas cooling and collapse begin in a DM minihalo (i.e., $z$ and $\Mvir$).
For the same cosmological density fluctuation, a stronger SV influence leads to a delay in the formation epoch and an increase in the virial halo mass, which results in the move from the bottom-right toward the top-left in the $z$-$\Mvir$ diagram (Figure~\ref{fig:z-M_each}).
As an example, consider I01, the density fluctuation that undergoes the earliest halo growth among those examined in this study.
Since the SV amplitude decreases with time as $\vsv \propto (1+z)$, the earliest-forming DM halo (I01) experiences the strongest SV effect of all models (I01-I20).
For I01, going from $\vsvsigma=0$ to $3$ decreases $z$ by $z^{'}-z = dz = 13.47$ and increases $\Mvir$ by a factor of $\Mvir^{'}/\Mvir = \Delta \Mvir = 173$, the largest difference in our sample.
While the gas cloud within the DM minihalo struggles to collapse, the large-scale structure around the halo continues to grow, forming more substantial filaments and knots (Figure~\ref{fig:2dmap}).
As a result, by the time the primordial gas cloud finally begins to contract, the distribution of surrounding matter differs significantly from that in the no-SV case (I01V00).
We also confirm that during the delayed formation epoch, the ongoing formation and mergers of minihalos contribute to halo growth.

Next, we examine the response of the DM halo formation from 20 primordial density fluctuations to six different SV values on $z$-$\Mvir$ diagram (Figure~\ref{fig:z-M_each}).
Starting with the smallest SV amplitude in our parameter set (V10 with $\vsvsigma=1$), we immediately see a large shift in the $z$-$\Mvir$ diagram compared to the no-SV model (V00).
Averaging over all models with this same SV amplitude (V10 in Table~\ref{table:average}) and comparing them to the no-SV case (V00), we find that increasing SV from $\vsvsigma=0$ to 1 changes the formation redshift by $dz = 2.38$ and increases $\Mvir$ by a factor of $\Delta \Mvir \sim 2.46$.
Because $\vsvsigma=1$ is close to the most probable SV value in the universe, $\vsvsigma \sim 0.8$ \citep{Tseliakhovich2010}, this result highlights the importance of accounting for SV in modelling the overall first star formation process.
As we increase SV further, both $dz$ and $\Delta \Mvir$ generally grow larger, reflecting stronger SV, induced delays and mass enhancements in halo formation.
A subset of models even surpasses $\Tvir=8000$\,K, where atomic hydrogen cooling becomes relevant, though ultimately all halos in our sample rely on H$_{2}$-cooling during collapse (see Section~\ref{sec:res:jeans}).
Interestingly, for $\vsvsigma=2.5-3$, we discover an inverse trend: lower-redshift collapses occur at lower halo masses, contrary to the behaviour in models with lower $\vsvsigma$ (coloured areas in Figure~\ref{fig:z-M_each}).
This finding suggests a new dependence on SV, indicating that the slope of the critical halo mass versus redshift relation may change sign at high SV amplitudes.

The effects of SV on halo properties also vary with the formation epoch.
Halos that form at higher redshifts experience larger changes in both redshift and virial mass ($z$ and $\Mvir$) when SV is increased, compared to halos forming at lower redshifts (Figure~\ref{fig:z-M_each}).
As the amplitude of SV decreases over time according to $\vsv \propto (1+z)$, even in regions with initially high SV, the influence of SV becomes weaker if the magnitude of density fluctuations is small and structure formation delays.
To perform a quantitative comparison, we classify halos into three groups based on their formation epoch ({\it High}, {\it Middle}, and {\it Low}) and average their physical properties (Table~\ref{table:average}).
Figure~\ref{fig:z-M_average} reveals that the distance between the averaged properties on the $z$-$\Mvir$ diagram for the {\it High} group is greater than that for the {\it Low} group.
This indicates that the impact of SV on the $z$-$\Mvir$ relation is stronger for the {\it High} group than for the {\it Low} group.
When we compare the effect of streaming velocity on the average $\Mvir$ across the three groups (by comparing same symbols in Figure~\ref{fig:z-M_average}), we observe that in the absence of SV ($\vsvsigma=0$), the change from the {\it Low} to {\it Middle} group follows the typical redshift dependence of virial mass (as given by Equation~\ref{eq:Mvir}); however, the transition from the Middle to High group deviates from this trend.
At higher redshifts and lower $\Mvir$, the effects of accretion and merger-induced dynamical heating become relatively more pronounced, delaying cloud collapse and increasing $\Mvir$.
On the other hand, at $\vsvsigma=1-2$, the virial mass remains approximately constant among 3 groups, whereas at $\vsvsigma=2.5-3$, {\it Low} group collapses at lower $\Mvir$ compared to {\it High} group.
When modelling the effect of SV on the critical halo mass, the formation epoch must be accounted for by adjusting the influence of SV accordingly.

Figure~\ref{fig:Vsv-HaloCloud} summarizes the average physical quantities for all models and each group as a function of SV magnitude.
{\it All} in the main panels represents the average values of physical quantities for each SV, demonstrating that the average values depend on the SV magnitude.
The sub-panels show the distribution of each physical quantity for each SV, confirming that SV determines not only the mean values but also the distribution shapes and variances.
Returning to the main panels, the grey lines in the background illustrate the SV dependence of physical quantities for each model (I10-I20), and their averaged values are classified into three groups, {\it High}, {\it Middle}, {\it Low}.
Figure~\ref{fig:Vsv-HaloCloud}(a) and (b) show the dependence of $z$ and $\Mvir$ discussed above.
Figure~\ref{fig:Vsv-HaloCloud}(c) shows the baryon fraction, $\fbaryon = M_{\rm b}/(M_{\rm b} + M_{\rm DM}) = M_{\rm b}/\Mvir$.
$\fbaryon$ within the DM halos is generally below the cosmic mean, $\Omega_{\rm b}/\Omega_{\rm m}=0.15484$ (horizontal dotted line), and decreases with increasing $\vsvsigma$.
This aligns with the known tendency of SV to inhibit baryon accretion into halos.
An exception is the {\it High} group at $\vsvsigma = 2-3$, where $\fbaryon$ increases due to the large SV at high redshift.
In these models, the baryon density fluctuations that originally tracked the DM fluctuations at the cosmic recombination have escaped before DM halo formation.
Consequently, the gas that subsequently collapses inside the DM halo originates from different regions at the cosmic recombination era and is later accreted due to the streaming velocity.
Although gas flowing from outside the halo initially overshoots because of the high SV, it is eventually captured by the deep gravitational potential of massive DM halos, resulting in an enhanced $\fbaryon$.\footnote{The SIGO scenario shows that primordial baryon density fluctuations escape from dark matter density fluctuations, resulting in baryon-dominated objects that later collapse and become globular cluster progenitors.
The origins of collapsing baryons for our models and SIGO scenario differ from our results, so the two objects could coexist under higher SV.
However, we find no SIGO in our simulations.
One possible explanation for why our simulations do not capture SIGOs is that baryon collapse within the DM minihalo occurs so early that the simulation is terminated before SIGOs can form outside the halo, or that the baryons which would seed SIGO formation are advected out of the computational domain by the streaming velocity.}

%%%%%%%%%%%%%%%%%%%%%%%%%%%%%%%%%%%%%%%%%%%%%%%%%%
\begin{figure*}
\begin{center}
    \includegraphics[width=1.0\linewidth]{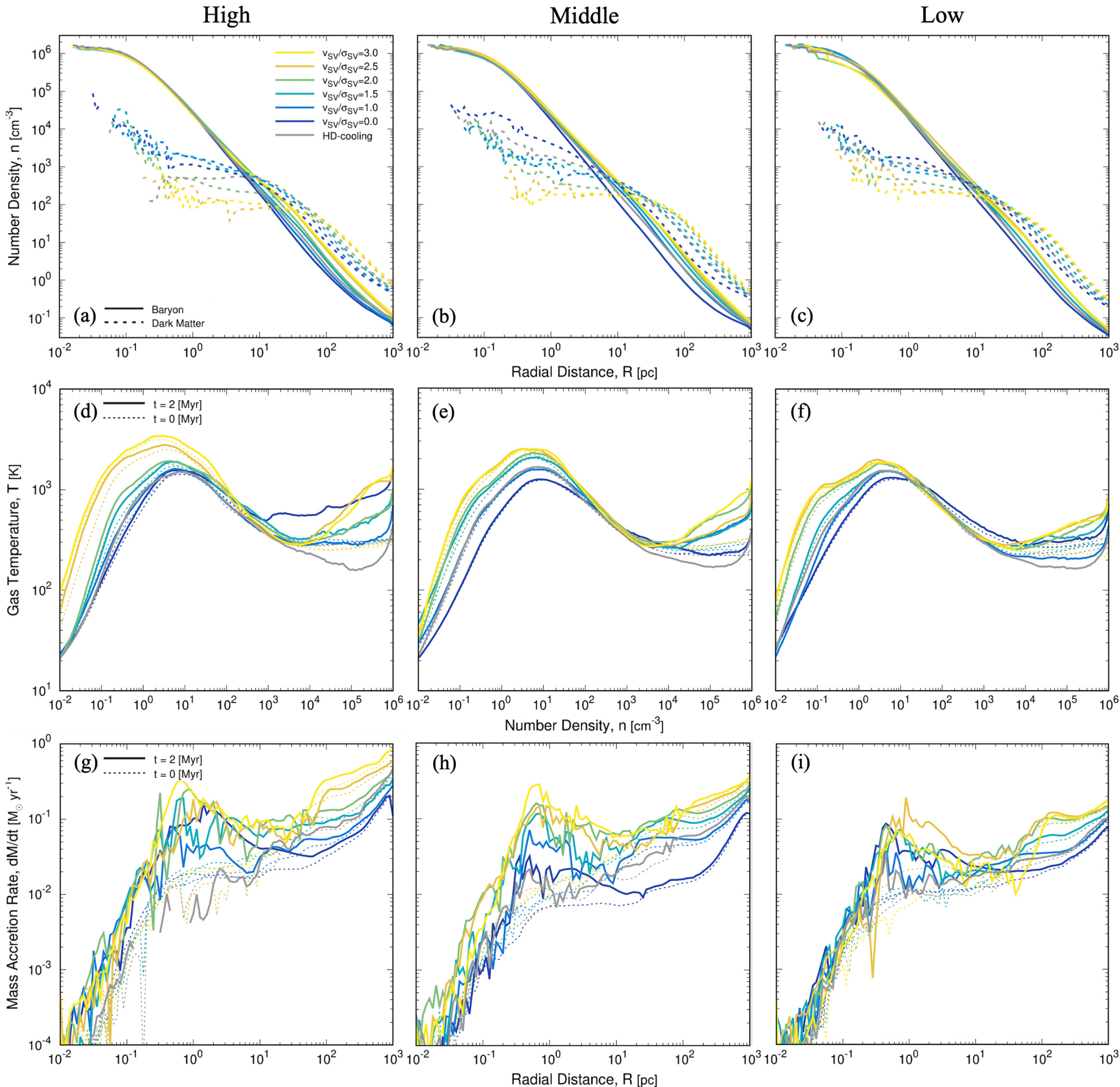}
\end{center}
\caption{
Profiles averaged across models belonging to three groups ({\it High}, {\it Middle}, and {\it Low}) for the same initial streaming velocity.
Panels: (a-c) radial profile of baryon (gas; solid lines) and dark matter (dashed) density when $\tth = 0$\,Myr, (d-f) density profile of gas temperature when $\tth = 0$\,Myr (dashed) and $\tth = 2$\,Myr (solid), and (g-i) radial profile of gas mass accretion rate when $\tth = 0$\,Myr (dashed) and $\tth = 2$\,Myr (solid).
The coloured lines show averaged profiles for each initial streaming velocity (V00-30), whereas the grey lines represent the HD-cooling models (see Table~\ref{table:each}).
}
\label{fig:R-Cloud}
\end{figure*}
%%%%%%%%%%%%%%%%%%%%%%%%%%%%%%%%%%%%%%%%%%%%%%%%%%

%%%%%%%%%%%%%%%%%%%%%%%%%%%%%%%%%%%%%%%%%%%%%%%%%%
\subsubsection{Exceptional correlation}
\label{sec:res:virial:exception}
%%%%%%%%%%%%%%%%%%%%%%%%%%%%%%%%%%%%%%%%%%%%%%%%%%

As confirmed thus far, generally, as the magnitude of SV in the halo formation region increases from $\vsvsigma=0$ to 3, halo formation delays and halo mass increases (lower redshift $z$ and higher halo mass $\Mvir$).
However, a small subset of models ($\sim\!10\%$) shows a different trend, for example, where collapse occurs at higher $z$ and lower $\Mvir$.
We classify the relevant models into three exceptional categories ({\it E1}, {\it E2}, and {\it E3}) according to the increase or decrease in $z$ and $\Mvir$ compared to models with smaller SV (Column~7 in Table\ref{table:each}).
\begin{itemize}
    \item {\it E1} as earlier formation of lighter halos (increase of $z$ and decrease of $\Mvir$): 10 models, I03V15, I04V30, I07V20, I07V30, I09V10, I15V20, I17V20, I18V10, I20V10, I20V30.
    \item {\it E2} as earlier formation of heavier halos (increases of both $z$ and $\Mvir$): 2 models, I15V30, I20V25.
    \item {\it E3} as later formation of lighter halos (decreases of both $z$ and $\Mvir$): 2 models, I16V30, I19V10.
\end{itemize}
While most models (106 models) follow the general trend of delayed formation and increased halo mass with higher SV, a subset of models (14 models) deviates from this relation.
These deviations suggest that SV's influence on halo formation is not uniform across all density fluctuations and may depend on additional factors such as local density environments or merger histories.

%%%%%%%%%%%%%%%%%%%%%%%%%%%%%%%%%%%%%%%%%%%%%%%%%%
\subsection{Jeans gas cloud}
\label{sec:res:jeans}
%%%%%%%%%%%%%%%%%%%%%%%%%%%%%%%%%%%%%%%%%%%%%%%%%%

By examining the properties of DM halos, we can investigate the conditions under which the first stars form, specifically, when and where they form.
Conversely, by analyzing gas cloud properties, we can explore the mass distribution of the first stars.
As discussed in Section~\ref{sec:res:virial}, the magnitude of SV alters the physical properties ($z$, $\Mvir$, and $\fbaryon$) of the DM halos where primordial stars form.
Consequently, the physical properties of the star-forming gas clouds within these halos are also influenced by SV.
In particular, the physical quantities at the Jeans scale of the star-forming gas clouds critically determine the star formation process of primordial stars.
To investigate how the magnitude of SV affects gas clouds' physical properties, we plotted the radial profiles of density, temperature, and accretion rate as functions of radius from the dense core (Figure~\ref{fig:R-Cloud}).
To determine whether the SV dependence of gas cloud properties varies with the halo's formation redshift and mass, we created radial profiles for each of the three classified groups: {\it High}, {\it Middle}, and {\it Low}.

Figure~\ref{fig:R-Cloud}(a-c) shows the gas (baryon) and DM density distributions at the end of the collapse phase ($\tth=0$\,yr).
Higher SV typically leads to more massive halos and larger radii of the DM halo ($\overline{\Mvir} \sim 10^6-10^7\,\msun$ and $\overline{\Rvir} \sim 100-500$\,pc as Table~\ref{table:average}).
The power-law exponent of the DM density distribution decreases around the virial radius $\Rvir$, resulting in a cuspy density profile.
Gas densities surpass DM densities at some radius ($R\sim\!10$\,pc, $n\sim10^2-10^3\,\cc$), but the exact crossing point depends on SV and halo class.
High SV tends to flatten the DM cusp into a core-like structure, shifting the gas-DM density crossing point outward and to lower densities.
This can facilitate the formation of large-scale sheets and filaments (Figure~\ref{fig:2dmap}) since gas collapses without being tightly bound by a steep DM potential well.

Figure~\ref{fig:R-Cloud}(d-f) shows the temperature distribution of the gas clouds at the onset of the accretion phase ($\tth=0$\,yr) and at the end of the simulation ($\tth=2$\,Myr).
Gas clouds are accumulated by the halo's gravity and undergo adiabatic compression as $n$ increases ($n \to 1\,\cc$).
At this stage, gas temperatures increase with SV magnitude, corresponding to increases in virial mass and virial temperature.
Furthermore, the {\it High} class is more strongly affected by SV compared to the {\it Middle} and {\it Low} classes.
As gas density increases ($n>1\,\cc$), the gas cools and contracts via H$_2$-cooling.
During the collapse phase (until $\tth=0$\,yr), the gas temperature does not show a clear dependence on SV.
During the accretion phase (e.g., $\tth=2$\,Myr), on the other hand, the gas temperature of the inner, dense region the temperature of dense gas inside the cloud shows an SV dependence, generally tending to increase with higher SV values.\footnote{The temperature of the {\it High} class with $\vsvsigma=0$ is higher than other averaged profiles with $\vsvsigma>0$ (Figure~\ref{fig:R-Cloud}d). This is because two models (I08V00 and I09V00) among the five models that were averaged have higher temperatures than the others.}
In some models, HD-cooling becomes effective, causing the gas temperature to drop below the temperature plateau associated with H$_2$-cooling (see Section~\ref{sec:res:jeans:HD}).

Figure~\ref{fig:R-Cloud}(g-i) shows the mass accretion rate of the gas clouds ($dM/dt=4 \pi r^2 n \vrad$).
Higher SV generally leads to increased accretion rates at the outer scales of the gas cloud (minihalo scale, $R>10$\,pc).
Conversely, at inner scales, at $\tth=0$\,yr, the influence of SV is minimal.
However, by $\tth=2$\,Myr, the accretion rates around the dense core increase, with slightly higher accretion rates observed for larger SV values.
When dividing the models into three classes, the overall accretion rate decreases in the order of {\it High}, {\it Middle}, and {\it Low}.
This suggests that gas clouds forming from high-redshift, high-mass halos, those with denser density fluctuations influenced by high SV, exhibit higher accretion rates.

%%%%%%%%%%%%%%%%%%%%%%%%%%%%%%%%%%%%%%%%%%%%%%%%%%
\begin{figure}
\begin{center}
\includegraphics[width=1.0\linewidth]{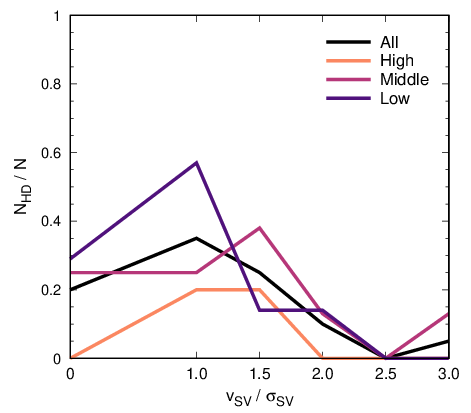}
\end{center}
\caption{
Initial streaming velocity dependence of the proportion of the HD-cooling models that meet the abundance ratio criterion $f_{\rm HD}/f_{\rm H_2} \geq 10^{-3}$ at the end of the calculation $\tth=2$\,Myr.
The black line shows the results averaged over all models ({\it All}), whereas the coloured lines show the averaged results for three classified groups ({\it High}, {\it Middle}, and {\it Low}).
}
\label{fig:Vsv-NhdN}
\end{figure}
%%%%%%%%%%%%%%%%%%%%%%%%%%%%%%%%%%%%%%%%%%%%%%%%%%

%%%%%%%%%%%%%%%%%%%%%%%%%%%%%%%%%%%%%%%%%%%%%%%%%%
\begin{figure*}
\begin{center}
\includegraphics[width=1.0\linewidth]{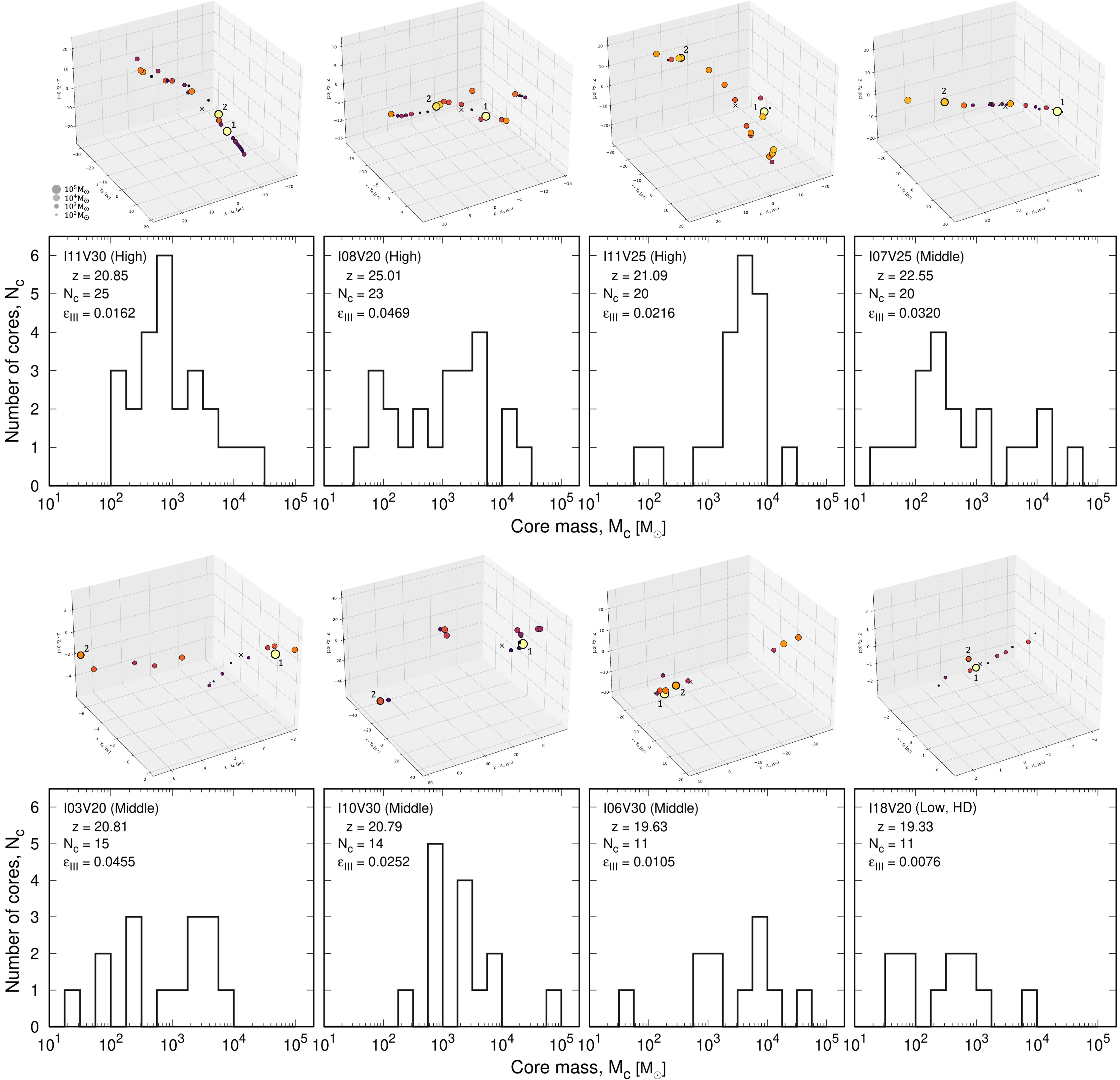}
\end{center}
\caption{
Core mass distribution for 8 models with large $\Ncore (\geq 10)$.
We also plot the spatial distribution of the cores for each system.
The larger the core mass, the larger the symbol's radius and the redder the symbol's colour.
We have distinguished the primary core, which has the largest mass in the system, and the secondary core, which has the second largest core mass, by indexing them as ``1'' and ``2''.
}
\label{fig:8cluster}
\end{figure*}
%%%%%%%%%%%%%%%%%%%%%%%%%%%%%%%%%%%%%%%%%%%%%%%%%%

%%%%%%%%%%%%%%%%%%%%%%%%%%%%%%%%%%%%%%%%%%%%%%%%%%
\subsubsection{HD-cooling}
\label{sec:res:jeans:HD}
%%%%%%%%%%%%%%%%%%%%%%%%%%%%%%%%%%%%%%%%%%%%%%%%%%

Besides H$_2$, hydrogen deuteride (HD) is also vital for first star formation \citep[e.g.,][]{Hirano2014}.
We determined whether a model exhibits effective HD-cooling by setting the condition that the chemical abundance ratio satisfies $f_{\rm HD}/f_{\rm H_2} \geq 10^{-3}$ at the end of the calculation, $\tth=2$\,Myr, (column 8 in Table~\ref{table:each}).
Nineteen models exhibit effective HD-cooling, which can further reduce gas temperatures below the H$_2$-cooling floor.
Such conditions lower the Jeans mass, potentially influencing the number and mass of dense cores formed inside.
As shown in Figure~\ref{fig:R-Cloud}(d-i), the average temperature of HD-cooling models at $\tth=2$\,Myr is lower, and the accretion rate is reduced compared to the average of other models due to cooling via H$_2$.

Table~\ref{table:average} and Figure~\ref{fig:Vsv-NhdN} summarize the fraction of HD-cooling models in each SV class.
We find that HD-cooling becomes ineffective for halos with $\vsvsigma \gtrsim 2$.
In these systems, the higher virial temperature and reduced baryon fraction likely prevent the gas from reaching the lower temperatures required for significant HD formation.
By contrast, halos with weaker SV retain higher gas densities at moderate temperatures (see solid lines in Figure~\ref{fig:R-Cloud}(d-f)), allowing HD to form and cool the gas below the temperature floor due to the H$_2$-cooling.
For example, HD-cooling becomes effective for about 60\% of L10 models, which were formed later with $\vsvsigma = 1$.
The average redshift of this group is $\overline{z}=20.95$, a period when the star formation rate density (SFRD) of the first stars in the early universe was still rising.
This synchronization suggests that the epoch of the most active first star formation coincides with the period when HD-cooling clouds account for more than half of all models.
Consequently, low-mass first stars formed under the influence of HD-cooling may constitute a significant proportion during this epoch.\footnote{\cite{Lenoble2024} recently discussed the relationship between HD-cooling and slow contraction by halo spin. Furthermore, the impact of external photo-dissociation, which is not considered in this study, on HD-cooling clouds is discussed in \citep{Nishijima2024}.}
This suggests that HD-cooling might shift the core mass function towards lower masses compared to the canonical first star formation scenario without considering SV.

%%%%%%%%%%%%%%%%%%%%%%%%%%%%%%%%%%%%%%%%%%%%%%%%%%
\subsection{Dense core}
\label{sec:res:CMF}
%%%%%%%%%%%%%%%%%%%%%%%%%%%%%%%%%%%%%%%%%%%%%%%%%%

After the initial gravitational collapse of the first gas cloud within each model ($n=\nth$), we continued the simulation for an additional 2\,Myr.
During this period, the gravitational contraction of gas within the minihalo progresses.
At this stage, the initially formed dense core where $n \geq \nth$ not only grows in mass through accretion but also, in some models, other regions of the primordial gas cloud undergo gravitational contraction, leading to the formation of additional dense cores (cloud-scale fragmentation).
We analyzed the final number of dense cores ($\Ncore$) and their masses ($\Mcore$) at the end of the simulation ($\tth=2$\,Myr) for each model (Table~\ref{table:each}).
Assuming that each dense core forms one first star, $\Ncore$ corresponds to the number of first stars, and $\Mcore$ represents the upper mass limit of these stars.\footnote{However, if disk-scale fragmentation occurs, a single dense core may host multiple first stars \citep[e.g.,][]{Susa2019, Sugimura2023}. In such cases, $\Mcore$ can be considered the upper limit of the total mass of multiple first stars.}

First, we introduce the models that exhibited particularly high fragmentation counts.
Figure~\ref{fig:8cluster} summarizes eight models in which the number of dense cores reached $\Ncore \geq 10$ by the end of the simulation.
Arranged in order of $\Ncore$ from top left to right and then bottom left to right, the top panels display the 3D distribution of dense cores.
Dense cores are formed along filaments with lengths of several tens of parsecs (up to 50\,pc in model I11V30).
Observing the temporal evolution, we confirmed that dense cores move along the filaments and approach each other.
We also observed cases where multiple dense cores merged within the 2\,Myr.
In models I10V30 and I06V30, dense core clusters independently form in distant regions of the filament.
The bottom panels show the mass distribution of dense cores, which spans a wide range from $\Mcore = 10 - 10^5\,\msun$.
Notably, multiple massive dense cores with $\Mcore \geq 10^4\,\msun$ are formed.
These massive cores acquire mass through the accretion of surrounding gas and mergers with other cores.
Each is thought to host a massive first star.
Additionally, we observed massive dense cores on the same filament approaching each other.

From Table~\ref{table:average}, we can examine the dependence of the number and mass of dense cores on the magnitude of SV.
The right-side sub-panels represent the distribution for each SV magnitude, illustrating SV's impact on each physical quantity.
For $\Ncore$ (Figure~\ref{fig:Vsv-HaloCloud}e), we find a wide diversity in the number of cores formed by 2\,Myr. 
Some halos remain monolithic, producing only a single core, while others fragment extensively, hosting more than 10 cores along elongated filaments.
Multiple fragmentation events are particularly common in halos that collapse at intermediate redshifts and have moderate SV values, conditions in which large-scale filamentary structures can grow.
Comparisons of $\Ncore$, total core mass ($\Mcoretot$; Figure~\ref{fig:Vsv-HaloCloud}f), and the primary core mass ($\Mcorefirst$; Figure~\ref{fig:Vsv-HaloCloud}g) reveal that while SV strongly influences the number of cores, the mass of cores remains broadly similar across the {\it High}, {\it Middle}, and {\it Low} classes.
Figure~\ref{fig:Vsv-HaloCloud}(h) plots $\qcore$, the mass ratio of the primary and secondary cores.
As SV increases, $\qcore$ slightly decreases, generally around $\qcore \sim 0.3$.
Large filaments sometimes host multiple cores with substantial masses, potentially serving as progenitors of star clusters or massive BH binary formation sites.

\begin{figure}
\begin{center}
\includegraphics[width=1.0\linewidth]{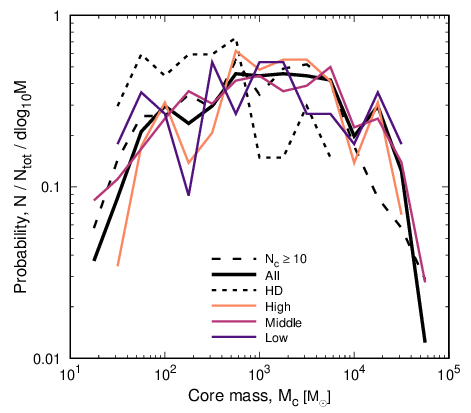}
\end{center}
\caption{
Mass function of dense cores when $\tth = 2$\,Myr normalized by $d\log_{10}M_{\rm c} = 0.25$.
We have combined the functions of models with $\Ncore \geq 3$ for five classes: {\it All}, {\it HD}, {\it High}, {\it Middle}, and {\it Low}.
The dashed line represents the combined function for models with large $\Ncore (\geq 10)$.
}
\label{fig:CMF}
\end{figure}

Figure~\ref{fig:CMF} summarizes the core mass function (CMF) of models with multiple dense core formations ($\Ncore \geq 3$) at the end of the simulation.
The overall average (black solid line) exhibits a nearly flat slope across $\Mcore = 5\times(10^2 - 10^3)\,\msun$.
Additionally, a sub-peak appears at $\Mcore = 2\times10^4\,\msun$, indicating that primary cores have grown to massive masses.
The formation epochs ({\it High}, {\it Middle}, and {\it Low}) do not significantly impact the CMF.
SV’s main role is to set the fragmentation mode (single versus multiple cores) rather than imposing a fundamentally different CMF shape.
In models that underwent extensive fragmentation with $\Ncore \geq 10$ (dashed line in Figure~\ref{fig:8cluster}), the sub-peak at $\sim 2\times10^4\,\msun$ in the CMF disappears.
In HD-cooling models ({\it HD}; dotted line), the distribution shifts towards the lower mass side compared to the average mass function, and massive cores with $\Mcore > 10^4\,\msun$ do not form.

%%%%%%%%%%%%%%%%%%%%%%%%%%%%%%%%%%%%%%%%%%%%%%%%%%
%%%%%%%%%%%%%%%%%%%%%%%%%%%%%%%%%%%%%%%%%%%%%%%%%%

%%%%%%%%%%%%%%%%%%%%%%%%%%%%%%%%%%%%%%%%%%%%%%%%%%
\section{Formation criterion of the first star clusters}
\label{sec:dis}
%%%%%%%%%%%%%%%%%%%%%%%%%%%%%%%%%%%%%%%%%%%%%%%%%%

A set of simulation results shows that SV governs a cascade of changes, from the halo formation epoch and mass to the thermodynamics and accretion dynamics of the Jeans-scale gas cloud and finally to the number and mass distribution of dense cores.
While the overall fragmentation behaviour depends on multiple factors (e.g., halo density fluctuations, cooling physics, etc.), SV emerges as a key parameter controlling whether the first star-forming regions produce a solitary massive star or a cluster of proto-stellar cores.

The simulation set presents that the halo mass and number of cores vary with the initial SV value.
This section formulates and qualitatively understands the dependences on the initial SV value.

%%%%%%%%%%%%%%%%%%%%%%%%%%%%%%%%%%%%%%%%%%%%%%%%%%
\subsection{Critical halo mass}
\label{sec:dis:Mcrit}
%%%%%%%%%%%%%%%%%%%%%%%%%%%%%%%%%%%%%%%%%%%%%%%%%%

Figure~\ref{fig:z-M_each} shows how the host halo mass at the onset of cloud collapse (the so-called critical halo mass, $\Mcrit$) depends on SV.
This critical halo mass is a key parameter for semi-analytical modelling \citep[e.g.,][]{Feathers2024}.
Using Kernel Density Estimation (KDE; coloured areas) based on our sample of halos, 
We confirm that the virial mass, $\Mvir$, varies not only with redshift but also systematically with SV.
In general, $\Mvir$ increases as SV increases, which is consistent with previous studies\citep[e.g.,][]{, Schauer2021SIGO, Kulkarni2021}.
In addition, we find that the redshift dependence of $\Mvir$ changes with SV; for higher SV values, $\Mvir$ tends to increase more steeply at higher redshifts, resulting in an inverse correlation compared to the typical trend.
Motivated by these results, we attempt to fit our data using the same functional form employed by \cite{Hirano2018}, who examined only three models with different SV for a single halo.

%%%%%%%%%%%%%%%%%%%%%%%%%%%%%%%%%%%%%%%%%%%%%%%%%%
\begin{figure}
\begin{center}
\includegraphics[width=1.0\linewidth]{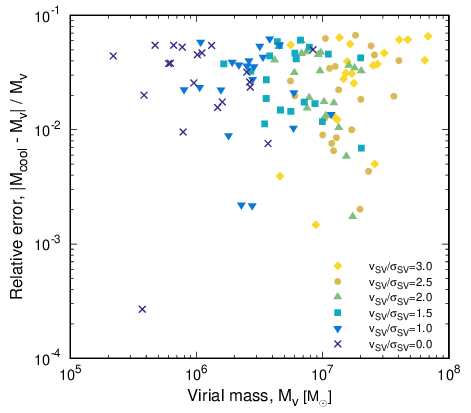}
\end{center}
\caption{
Relative error of the cooling threshold mass to the halo mass.
We assign the value obtained from the analysis of the simulation data to the circular velocity in Equation~\ref{eq:Mcool}.
}
\label{fig:Mcool}
\end{figure}
%%%%%%%%%%%%%%%%%%%%%%%%%%%%%%%%%%%%%%%%%%%%%%%%%%

%%%%%%%%%%%%%%%%%%%%%%%%%%%%%%%%%%%%%%%%%%%%%%%%%%
\begin{figure*}
\begin{center}
\includegraphics[width=1.0\columnwidth]{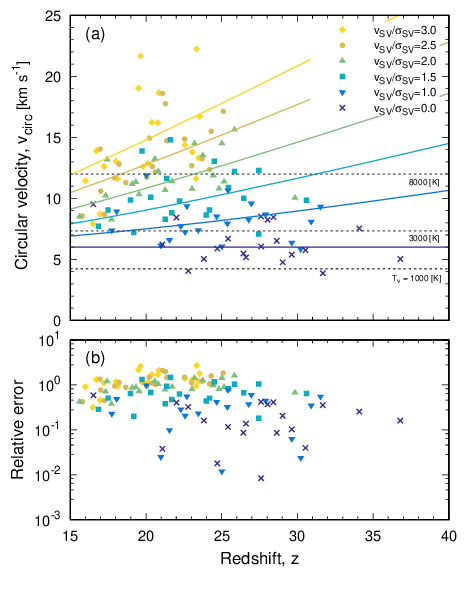}
\includegraphics[width=1.0\columnwidth]{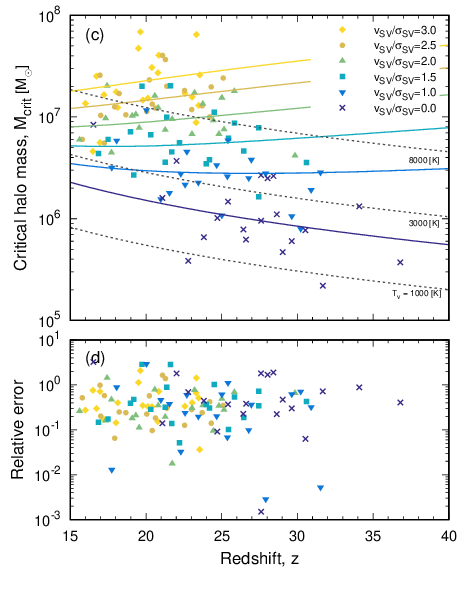}
\end{center}
\caption{
Fitting functions of circular velocity (Equation~\ref{eq:vcirc_Fialkov2012}, left panels) and critical halo mass (as Equation~\ref{eq:Mcool}, right) with $v_{\rm circ,0} = 6.0\,\kms$, $\alpha = 7.8$.
The top panels compare the fitting functions and simulation results.
The bottom panels show the relative error of the fitting functions to the simulation results.
The dashed lines in panels (a) and (c) indicate the circular velocities and virial masses for three different virial temperatures, $\Tvir = 1000$, $3000$, and $8000$\,K (Equations~\ref{eq:Mvir} and \ref{eq:Mcool}).
}
\label{fig:z-Vcirc}
\end{figure*}
%%%%%%%%%%%%%%%%%%%%%%%%%%%%%%%%%%%%%%%%%%%%%%%%%%

%%%%%%%%%%%%%%%%%%%%%%%%%%%%%%%%%%%%%%%%%%%%%%%%%%
\begin{figure}
\begin{center}
\includegraphics[width=1.0\linewidth]{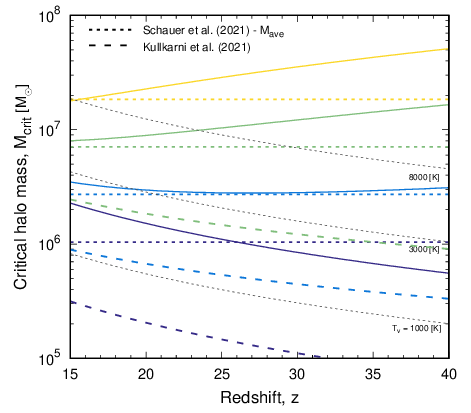}
\end{center}
\caption{
Comparison of fitting functions of the critical halo mass: (solid) Equations~\ref{eq:Mcool} and \ref{eq:vcirc_Fialkov2012} with $\vsvsigma=0$ 1, 2, and 3, (dotted) Equation~9 of \citet{Schauer2021SV+UV} with $\vsvsigma=0$, 1, 2, and 3, and (dashed) Table~2 of \citet{Kulkarni2021} with $\vsvsigma=0$, 1, and 2.
The thin dashed lines indicate the virial halo masses for $\Tvir=1000$, 3000, and 8000\,K.
}
\label{fig:z-Mcrit_comp}
\end{figure}
%%%%%%%%%%%%%%%%%%%%%%%%%%%%%%%%%%%%%%%%%%%%%%%%%%

%%%%%%%%%%%%%%%%%%%%%%%%%%%%%%%%%%%%%%%%%%%%%%%%%%
\begin{figure*}
\begin{center}
\includegraphics[width=1.0\linewidth]{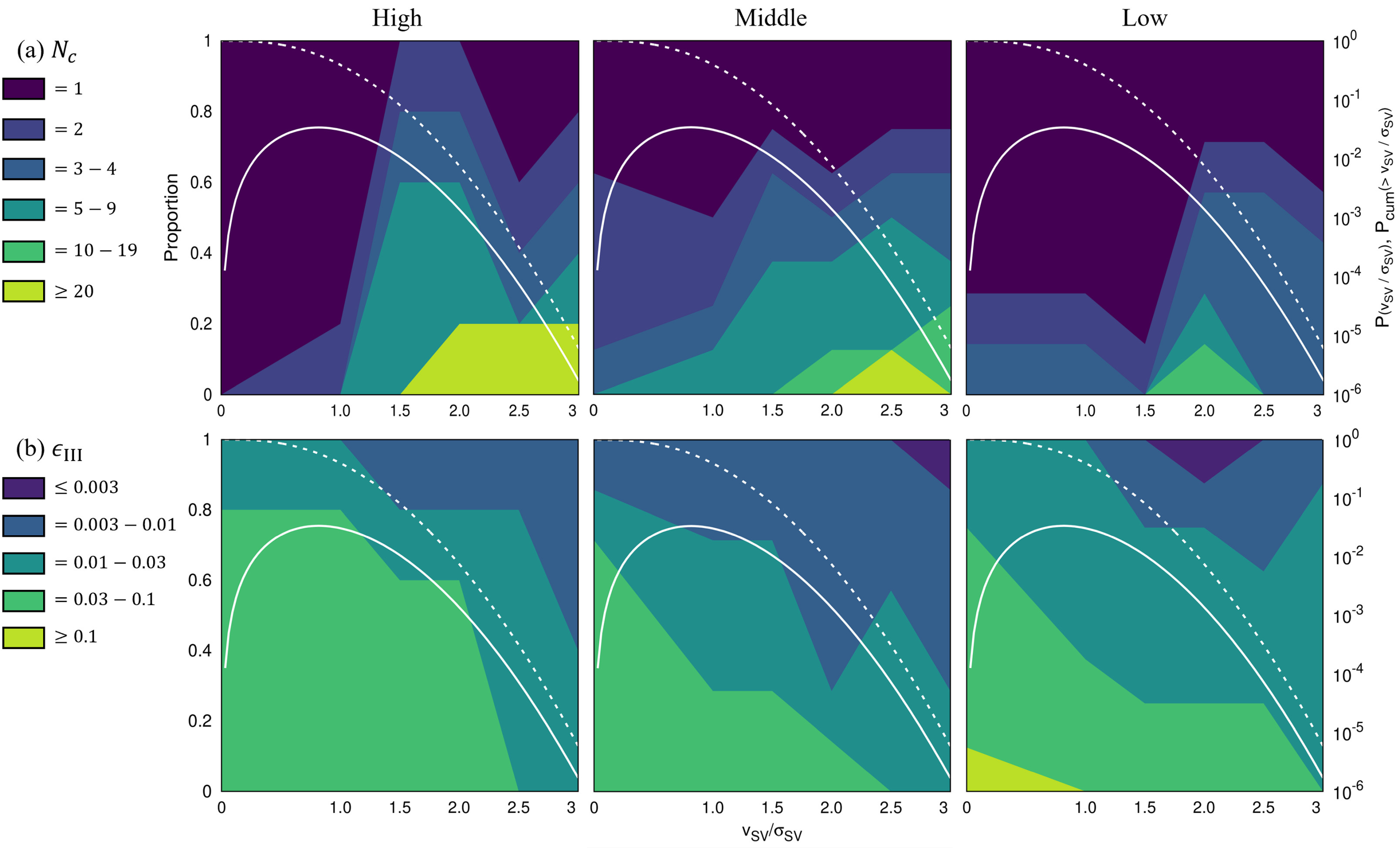}
\end{center}
\caption{
Dependence of the proportions of the number of cores ($\Ncore$) and mass conversion efficiency ($\epsIII$) on the initial streaming velocity for three classes ({\it High}, {\it Middle}, and {\it Low}).
The white lines show the probability distribution of the streaming velocity \citep[solid; Equation~13 in][]{Tseliakhovich2011} and the cumulative one (dashed).
}
\label{fig:Vsv-Nc_eqsIII}
\end{figure*}
%%%%%%%%%%%%%%%%%%%%%%%%%%%%%%%%%%%%%%%%%%%%%%%%%%

We define the critical halo mass, $\Mcrit$, as the virial mass $\Mvir$ at the onset of the gas collapse.
This threshold closely matches the so-called cooling threshold mass $\Mcool$:
%, defined at the characteristic temperature $\Tvir = 1000$\,K:
\begin{eqnarray}
    \Mcool = 1.2 \times 10^6\,\msun \left( \frac{\vcirc}{6\,\kms} \right)^3 \left( \frac{1+z}{25} \right)^{-3/2} \ ,
    \label{eq:Mcool}
\end{eqnarray}
where $\vcirc = \sqrt{G\Mvir/\Rvir}$ is the circular velocity, $G$ is the gravitational constant, and $\Mvir$ and $\Rvir$ are the virial mass and radius, respectively \citep[e.g.,][]{BarkanaLoeb2001}.

Before proceeding with the fitting, we verify that the halo dynamics characterized by $\vcirc$ can reliably represent the cooling threshold mass.
To do this, we compare $\Mcool$ to the measured $M_{\rm vir}$ for our 120 models and compute their relative differences (Figure~\ref{fig:Mcool}).
We find that, regardless of the SV value, the relative error is less than 10\% for all models.
Thus, if we can construct a suitable fitting function for $\vcirc$, we can reliably determine $M_{\rm crit}$.

\cite{Fialkov2012} proposed a formulation for the increase in the critical halo mass by introducing an ``effective'' velocity:
\begin{eqnarray}
    \vcirc = \sqrt{v_{\rm circ,0}^2 + [\alpha v_{\rm SV}(z)]^2} \ ,
    \label{eq:vcirc_Fialkov2012}
\end{eqnarray}
where $v_{\rm circ,0}$ and $\alpha$ are fitting parameters.
Using our sample, we re-fit these parameters and find $v_{\rm circ,0} = 6.0\,\kms$ and $\alpha = 7.8$, which best reproduce our results.\footnote{For comparison, \cite{Hirano2018} reported $v_{\rm circ,0} = 3.7\,\kms$ and $\alpha = 4.0$.}
Figure~\ref{fig:z-Vcirc} shows the resulting relations between $\vcirc$, $\Mcrit$, and $\Mvir$.
Although the overall trends are captured, panels (a) and (c) reveal significant scatter around the fitting lines.
Panels (b) and (d) show that relative errors can sometimes reach a factor of 3.

The key difference between the fitting functions obtained in this and previous works is that, as SV increases, the redshift dependence of the critical halo mass ($\Mcrit$) switches from increasing to decreasing with decreasing redshift.
Figure~\ref{fig:z-Mcrit_comp} compares our $\Mcrit$ model with those derived from earlier numerical simulations.
For $\vsvsigma = 0$, the redshift dependence of $\Mcrit$ is consistent with the results of \cite{Kulkarni2021}, and the overall increase in $\Mcrit$ with increasing SV is similarly in \cite{Schauer2021SIGO}.
The new feature is that $\Mcrit$ increases with redshift for higher $\vsvsigma$.
This means that the SV dependence of $\Mcrit$ in Equation~\ref{eq:Mcool}, where $\vcirc^3 \propto v_{\rm SV}^3 \propto (1+z)^3$, dominates over the redshift dependence $(1+z)^{-3/2}$ at high redshift.
The critical halo mass derived from analytical modeling qualitatively suggests a similar trend \citep{Nebrin2023}.
We consider that previous numerical works missed this trend due to insufficient samples of halos at higher redshifts.

This behaviour can explain the presence of models that far exceed the conventional upper mass limit for H-cooling halos ($\Tvir \simeq 8000$\,K).
Another possible interpretation is that at high redshifts, mergers of minihalos introduce additional complexity.
Simple temperature thresholds like $\Tvir=1000$\,K may not fully capture the evolutionary dynamics \citep[e.g., ``violent merger delay'' scenario][]{Inayoshi2018, Wise2019}.
Preliminary investigations into even higher-redshift, more massive halos suggest that a higher virial temperature threshold, such as $\Tvir=8000$\,K, might better explain these systems.

%%%%%%%%%%%%%%%%%%%%%%%%%%%%%%%%%%%%%%%%%%%%%%%%%%
\subsection{Formation criterion of the first star cluster}
\label{sec:dis:Ncore}
%%%%%%%%%%%%%%%%%%%%%%%%%%%%%%%%%%%%%%%%%%%%%%%%%%

To investigate the formation criterion of the first star cluster, we focus on the mass distribution of dense cores at 2\,Myr after the onset of the star-forming region, rather than directly counting the number of actual first stars.
Since these cores emerge from the fragmentation of large-scale filaments and sheets, the core mass distribution provides an upper-limit criterion for identifying potential massive first star clusters.

We next examine how the number of cores ($\Ncore$) depends on the baryonic streaming velocity (SV).
As shown in Figure~\ref{fig:Vsv-HaloCloud}(e), a clear transition occurs at around $\vsvsigma = 1.5$: systems with lower SV values produce only a single core, while those above this threshold form multiple cores.
This transition is also evident in the frequency distributions (see the inset panel), highlighting a change from single to multiple core formation regimes.
At $\vsvsigma \geq 1-1.5$, the thermal structure \citep[e.g., $T_{\rm max}/T_{\rm mix}$][]{Hirano2023PaperI} and density evolution produce more extensive sheets and filaments (Figure~\ref{fig:R-Cloud}).
The fragmentation of these massive filaments leads to the simultaneous formation of multiple cores along their length (Figure~\ref{fig:2dmap} and \ref{fig:CMF}).
We also find indications that HD-cooling can enhance fragmentation, particularly in cases without SV.

To quantify the parameter space for multiple core formation, we plot the $\Ncore$ distributions for each SV value in Figure~\ref{fig:Vsv-Nc_eqsIII}(a).
As discussed, at $\vsvsigma \geq 1.5$, we observe a marked increase in $\Ncore$.
Higher halo mass or redshift environments ({\it High} and {\it Middle}) also tend to produce more cores than {\it Low} ones.

We overlay the probability density function of $\vsv$ onto our results. Around the cosmic mean SV amplitude of $\vsvsigma \approx 0.8$, the star-forming behaviour is essentially indistinguishable from the no-SV scenario.
Thus, in about 90\% of cases (with $\vsvsigma < 1.5$), the conventional 1 halo-1 cloud scenario remains valid.
Only a few per cent of cases (with $\vsvsigma \geq 1.5$) experience a 1 halo-multiple cloud regime, producing on average $\Ncore \sim 4$ (Table~\ref{table:average}), and occasionally $\Ncore \geq 10$ (Table~\ref{table:each}).

In high-redshift, high-$\Mvir$ regimes not examined in this study, previous work \citep{Hirano2017smbh} has shown that the collapse may lead to a single supermassive star.
In that upper-right region of the $z$-$\Mvir$ parameter space, filaments fail to fragment into multiple cores, instead experiencing rapid collapse that supports a high accretion rate onto a single object.
Consequently, the formation of first star clusters appears to be a phenomenon restricted to intermediate regions of the $z$-$\Mvir$ parameter space, where filament collapse is not as dominant, allowing multiple cores, and thus potential first star clusters, to form.

%%%%%%%%%%%%%%%%%%%%%%%%%%%%%%%%%%%%%%%%%%%%%%%%%%
\subsection{Mass conversion efficiency}
\label{sec:dis:efficiency}
%%%%%%%%%%%%%%%%%%%%%%%%%%%%%%%%%%%%%%%%%%%%%%%%%%

Finally, we examine the gas-to-core mass ratio, treating it as an upper limit on the gas-to-star conversion efficiency within the halo.
Figure~\ref{fig:Vsv-HaloCloud}(d).
This ratio decreases with increasing $\vsv$, (the mean values at each $\vsv$, decrease from 4.53\% to 0.89\%; Table~\ref{table:average}).
It also decreases as the amplitude of density fluctuations becomes larger, from {\it High} to {\it Middle} to {\it Low}.
The efficiency tends to decline at lower redshifts for models with larger $\vsv$, values.
Moreover, models with effective HD-cooling show the lowest efficiency among them.

Figure~\ref{fig:Vsv-Nc_eqsIII}(b).
We plot the distribution of this ratio for each $\vsv$, together with the probability distribution of $\vsv$, itself.
At the most common values, over 80\% of the systems have efficiencies of $1-10\%$.
Although this fraction decreases with increasing $\vsv$, it is only marginally evident within the current plot resolution.

We interpret these results as follows.
To maintain a high star formation efficiency, the entire large-scale, massive filament must undergo a nearly uniform density increase.
In reality, only the regions with strong density perturbations grow non-linearly first, so after 2\,Myr of evolution, only a portion of the filament has become a dense gas cloud.
Therefore, in scenarios where large-scale filaments form (i.e., at high $\vsv$) the star formation efficiency decreases.

%%%%%%%%%%%%%%%%%%%%%%%%%%%%%%%%%%%%%%%%%%%%%%%%%%
\section{Conclusion}
\label{sec:con}
%%%%%%%%%%%%%%%%%%%%%%%%%%%%%%%%%%%%%%%%%%%%%%%%%%

In summary, baryonic streaming velocities (SV) significantly affect the mass scale and fragmentation of the first star-forming halos.
Our large sample of 120 cosmological simulations demonstrates that:
\begin{itemize}
    \item \textbf{Higher SV} ($\vsvsigma \ge 1.5$) delays the first star formation, shifts into more massive halos, and promotes filamentary collapse that produces multiple dense cores within a single halo.
    \item \textbf{Lower SV} ($\vsvsigma \le 1.0$) also delays the first star formation but results in a negligible impact on the cloud evolution, typically forming only one or two cores.
    \item \textbf{HD-cooling} is more likely to be effective under the influence of moderate SV, especially in models that form in low-z, in more than half of them. This is the parameter space in which low-mass first stars form.
\end{itemize}
These results indicate the importance of including SV in theoretical models of the cosmic dawn.\footnote{For instance, the most probable streaming velocity amplitude across the universe is around $\vsvsigma \sim 0.8$ \citep{Tseliakhovich2011}, whereas an analysis specific to the Milky Way region reports a value of about $\vsvsigma \sim 1.75$ \citep{Uysal2023}.}
In particular, the formation of multiple dense cores in a single halo could yield the first star clusters, with implications for black hole seed formation and early chemical enrichment.
Future work that incorporates radiative feedback, magnetic fields, and disk-scale fragmentation will be crucial for building a comprehensive model of the first star formation.

%%%%%%%%%%%%%%%%%%%%%%%%%%%%%%%%%%%%%%%%%%%%%%%%%%
\section*{Acknowledgements}
%%%%%%%%%%%%%%%%%%%%%%%%%%%%%%%%%%%%%%%%%%%%%%%%%%

We thank Hyunbae Park for discussing the BTD approximation for the cosmological initial condition setting.
Numerical computations were carried out on Cray XC50 at CfCA in the National Astronomical Observatory of Japan and Yukawa-21 at YITP in Kyoto University.
Numerical analyses were, in part, carried out on the analysis servers at CfCA in the National Astronomical Observatory of Japan.
This work was supported by JSPS KAKENHI Grant Numbers JP21K13960 and JP21H01123 (S.H.).

%%%%%%%%%%%%%%%%%%%%%%%%%%%%%%%%%%%%%%%%%%%%%%%%%%
\section*{Data Availability}
%%%%%%%%%%%%%%%%%%%%%%%%%%%%%%%%%%%%%%%%%%%%%%%%%%

The data presented in Tables~\ref{table:average} and \ref{table:each} are publicly available at our GitHub repository: https://github.com/shingohirano-astro/FSC2.
The data underlying this article will be shared on reasonable request to the corresponding author.

%%%%%%%%%%%%%%%%%%%%%%%%%%%%%%%%%%%%%%%%%%%%%%%%%%
%%%%%%%%%%%%%%%%%%%% REFERENCES %%%%%%%%%%%%%%%%%%
%%%%%%%%%%%%%%%%%%%%%%%%%%%%%%%%%%%%%%%%%%%%%%%%%%

\bibliographystyle{mnras}
\bibliography{ms}

\begin{thebibliography}{}
\makeatletter
\relax
\def\mn@urlcharsother{\let\do\@makeother \do\$\do\&\do\#\do\^\do\_\do\%\do\~}
\def\mn@doi{\begingroup\mn@urlcharsother \@ifnextchar [ {\mn@doi@} {\mn@doi@[]}}
\def\mn@doi@[#1]#2{\def\@tempa{#1}\ifx\@tempa\@empty \href {http://dx.doi.org/#2} {doi:#2}\else \href {http://dx.doi.org/#2} {#1}\fi \endgroup}
\def\mn@eprint#1#2{\mn@eprint@#1:#2::\@nil}
\def\mn@eprint@arXiv#1{\href {http://arxiv.org/abs/#1} {{\tt arXiv:#1}}}
\def\mn@eprint@dblp#1{\href {http://dblp.uni-trier.de/rec/bibtex/#1.xml} {dblp:#1}}
\def\mn@eprint@#1:#2:#3:#4\@nil{\def\@tempa {#1}\def\@tempb {#2}\def\@tempc {#3}\ifx \@tempc \@empty \let \@tempc \@tempb \let \@tempb \@tempa \fi \ifx \@tempb \@empty \def\@tempb {arXiv}\fi \@ifundefined {mn@eprint@\@tempb}{\@tempb:\@tempc}{\expandafter \expandafter \csname mn@eprint@\@tempb\endcsname \expandafter{\@tempc}}}

\bibitem[\protect\citeauthoryear{{Aoki}, {Tominaga}, {Beers}, {Honda}  \& {Lee}}{{Aoki} et~al.}{2014}]{Aoki2014}
{Aoki} W.,  {Tominaga} N.,  {Beers} T.~C.,  {Honda} S.,   {Lee} Y.~S.,  2014, \mn@doi [Science] {10.1126/science.1252633}, \href {https://ui.adsabs.harvard.edu/abs/2014Sci...345..912A} {345, 912}

\bibitem[\protect\citeauthoryear{{Aoki}, {Matsuno}, {Honda}, {Ishigaki}, {Li}, {Suda}  \& {Kumar}}{{Aoki} et~al.}{2018}]{Aoki2018}
{Aoki} W.,  {Matsuno} T.,  {Honda} S.,  {Ishigaki} M.~N.,  {Li} H.,  {Suda} T.,   {Kumar} Y.~B.,  2018, \mn@doi [\pasj] {10.1093/pasj/psy092}, \href {https://ui.adsabs.harvard.edu/abs/2018PASJ...70...94A} {70, 94}

\bibitem[\protect\citeauthoryear{{Barkana} \& {Loeb}}{{Barkana} \& {Loeb}}{2001}]{BarkanaLoeb2001}
{Barkana} R.,  {Loeb} A.,  2001, \mn@doi [\physrep] {10.1016/S0370-1573(01)00019-9}, \href {https://ui.adsabs.harvard.edu/abs/2001PhR...349..125B} {349, 125}

\bibitem[\protect\citeauthoryear{{Bessell} et~al.,}{{Bessell} et~al.}{2015}]{Bessell2015}
{Bessell} M.~S.,  et~al., 2015, \mn@doi [\apjl] {10.1088/2041-8205/806/1/L16}, \href {https://ui.adsabs.harvard.edu/abs/2015ApJ...806L..16B} {806, L16}

\bibitem[\protect\citeauthoryear{{Caffau} et~al.,}{{Caffau} et~al.}{2011}]{Caffau2011}
{Caffau} E.,  et~al., 2011, \mn@doi [\nat] {10.1038/nature10377}, \href {https://ui.adsabs.harvard.edu/abs/2011Natur.477...67C} {477, 67}

\bibitem[\protect\citeauthoryear{{Chiou}, {Naoz}, {Marinacci}  \& {Vogelsberger}}{{Chiou} et~al.}{2018}]{Chiou2018}
{Chiou} Y.~S.,  {Naoz} S.,  {Marinacci} F.,   {Vogelsberger} M.,  2018, \mn@doi [\mnras] {10.1093/mnras/sty2480}, \href {https://ui.adsabs.harvard.edu/abs/2018MNRAS.481.3108C} {481, 3108}

\bibitem[\protect\citeauthoryear{{Chiou}, {Naoz}, {Burkhart}, {Marinacci}  \& {Vogelsberger}}{{Chiou} et~al.}{2019}]{Chiou2019}
{Chiou} Y.~S.,  {Naoz} S.,  {Burkhart} B.,  {Marinacci} F.,   {Vogelsberger} M.,  2019, \mn@doi [\apjl] {10.3847/2041-8213/ab263a}, \href {https://ui.adsabs.harvard.edu/abs/2019ApJ...878L..23C} {878, L23}

\bibitem[\protect\citeauthoryear{{Chiti} et~al.,}{{Chiti} et~al.}{2024}]{Chiti2024}
{Chiti} A.,  et~al., 2024, \mn@doi [Nature Astronomy] {10.1038/s41550-024-02223-w}, \href {https://ui.adsabs.harvard.edu/abs/2024NatAs...8..637C} {8, 637}

\bibitem[\protect\citeauthoryear{{Feathers}, {Kulkarni}, {Visbal}  \& {Hazlett}}{{Feathers} et~al.}{2024}]{Feathers2024}
{Feathers} C.~R.,  {Kulkarni} M.,  {Visbal} E.,   {Hazlett} R.,  2024, \mn@doi [\apj] {10.3847/1538-4357/ad1688}, \href {https://ui.adsabs.harvard.edu/abs/2024ApJ...962...62F} {962, 62}

\bibitem[\protect\citeauthoryear{{Fialkov}}{{Fialkov}}{2014}]{Fialkov2014}
{Fialkov} A.,  2014, \mn@doi [International Journal of Modern Physics D] {10.1142/S0218271814300171}, \href {https://ui.adsabs.harvard.edu/abs/2014IJMPD..2330017F} {23, 1430017}

\bibitem[\protect\citeauthoryear{{Fialkov}, {Barkana}, {Tseliakhovich}  \& {Hirata}}{{Fialkov} et~al.}{2012}]{Fialkov2012}
{Fialkov} A.,  {Barkana} R.,  {Tseliakhovich} D.,   {Hirata} C.~M.,  2012, \mn@doi [\mnras] {10.1111/j.1365-2966.2012.21318.x}, \href {https://ui.adsabs.harvard.edu/abs/2012MNRAS.424.1335F} {424, 1335}

\bibitem[\protect\citeauthoryear{{Greif}, {White}, {Klessen}  \& {Springel}}{{Greif} et~al.}{2011}]{Greif2011sv}
{Greif} T.~H.,  {White} S. D.~M.,  {Klessen} R.~S.,   {Springel} V.,  2011, \mn@doi [\apj] {10.1088/0004-637X/736/2/147}, \href {https://ui.adsabs.harvard.edu/abs/2011ApJ...736..147G} {736, 147}

\bibitem[\protect\citeauthoryear{{Hahn} \& {Abel}}{{Hahn} \& {Abel}}{2011}]{HahnAbel2011}
{Hahn} O.,  {Abel} T.,  2011, \mn@doi [\mnras] {10.1111/j.1365-2966.2011.18820.x}, \href {https://ui.adsabs.harvard.edu/abs/2011MNRAS.415.2101H} {415, 2101}

\bibitem[\protect\citeauthoryear{{Harikane} et~al.,}{{Harikane} et~al.}{2024}]{Harikane2024}
{Harikane} Y.,  et~al., 2024, \mn@doi [arXiv e-prints] {10.48550/arXiv.2406.18352}, \href {https://ui.adsabs.harvard.edu/abs/2024arXiv240618352H} {p. arXiv:2406.18352}

\bibitem[\protect\citeauthoryear{{Hirano} \& {Bromm}}{{Hirano} \& {Bromm}}{2017}]{HiranoBromm2017}
{Hirano} S.,  {Bromm} V.,  2017, \mn@doi [\mnras] {10.1093/mnras/stx1220}, \href {https://ui.adsabs.harvard.edu/abs/2017MNRAS.470..898H} {470, 898}

\bibitem[\protect\citeauthoryear{{Hirano}, {Hosokawa}, {Yoshida}, {Umeda}, {Omukai}, {Chiaki}  \& {Yorke}}{{Hirano} et~al.}{2014}]{Hirano2014}
{Hirano} S.,  {Hosokawa} T.,  {Yoshida} N.,  {Umeda} H.,  {Omukai} K.,  {Chiaki} G.,   {Yorke} H.~W.,  2014, \mn@doi [\apj] {10.1088/0004-637X/781/2/60}, \href {https://ui.adsabs.harvard.edu/abs/2014ApJ...781...60H} {781, 60}

\bibitem[\protect\citeauthoryear{{Hirano}, {Hosokawa}, {Yoshida}, {Omukai}  \& {Yorke}}{{Hirano} et~al.}{2015}]{Hirano2015}
{Hirano} S.,  {Hosokawa} T.,  {Yoshida} N.,  {Omukai} K.,   {Yorke} H.~W.,  2015, \mn@doi [\mnras] {10.1093/mnras/stv044}, \href {https://ui.adsabs.harvard.edu/abs/2015MNRAS.448..568H} {448, 568}

\bibitem[\protect\citeauthoryear{{Hirano}, {Hosokawa}, {Yoshida}  \& {Kuiper}}{{Hirano} et~al.}{2017}]{Hirano2017smbh}
{Hirano} S.,  {Hosokawa} T.,  {Yoshida} N.,   {Kuiper} R.,  2017, \mn@doi [Science] {10.1126/science.aai9119}, \href {https://ui.adsabs.harvard.edu/abs/2017Sci...357.1375H} {357, 1375}

\bibitem[\protect\citeauthoryear{{Hirano}, {Yoshida}, {Sakurai}  \& {Fujii}}{{Hirano} et~al.}{2018}]{Hirano2018}
{Hirano} S.,  {Yoshida} N.,  {Sakurai} Y.,   {Fujii} M.~S.,  2018, \mn@doi [\apj] {10.3847/1538-4357/aaaaba}, \href {https://ui.adsabs.harvard.edu/abs/2018ApJ...855...17H} {855, 17}

\bibitem[\protect\citeauthoryear{{Hirano}, {Shen}, {Nishijima}, {Sakai}  \& {Umeda}}{{Hirano} et~al.}{2023}]{Hirano2023PaperI}
{Hirano} S.,  {Shen} Y.,  {Nishijima} S.,  {Sakai} Y.,   {Umeda} H.,  2023, \mn@doi [\mnras] {10.1093/mnras/stad2693}, \href {https://ui.adsabs.harvard.edu/abs/2023MNRAS.525.5737H} {525, 5737}

\bibitem[\protect\citeauthoryear{{Inayoshi}, {Li}  \& {Haiman}}{{Inayoshi} et~al.}{2018}]{Inayoshi2018}
{Inayoshi} K.,  {Li} M.,   {Haiman} Z.,  2018, \mn@doi [\mnras] {10.1093/mnras/sty1720}, \href {https://ui.adsabs.harvard.edu/abs/2018MNRAS.479.4017I} {479, 4017}

\bibitem[\protect\citeauthoryear{{Keller} et~al.,}{{Keller} et~al.}{2014}]{Keller2014}
{Keller} S.~C.,  et~al., 2014, \mn@doi [\nat] {10.1038/nature12990}, \href {https://ui.adsabs.harvard.edu/abs/2014Natur.506..463K} {506, 463}

\bibitem[\protect\citeauthoryear{{Kitsionas} \& {Whitworth}}{{Kitsionas} \& {Whitworth}}{2002}]{KitsionasWhitworth2002}
{Kitsionas} S.,  {Whitworth} A.~P.,  2002, \mn@doi [\mnras] {10.1046/j.1365-8711.2002.05115.x}, \href {https://ui.adsabs.harvard.edu/abs/2002MNRAS.330..129K} {330, 129}

\bibitem[\protect\citeauthoryear{{Klessen} \& {Glover}}{{Klessen} \& {Glover}}{2023}]{KlessenGlover2023}
{Klessen} R.~S.,  {Glover} S. C.~O.,  2023, \mn@doi [\araa] {10.1146/annurev-astro-071221-053453}, \href {https://ui.adsabs.harvard.edu/abs/2023ARA&A..61...65K} {61, 65}

\bibitem[\protect\citeauthoryear{{Kulkarni}, {Visbal}  \& {Bryan}}{{Kulkarni} et~al.}{2021}]{Kulkarni2021}
{Kulkarni} M.,  {Visbal} E.,   {Bryan} G.~L.,  2021, \mn@doi [\apj] {10.3847/1538-4357/ac08a3}, \href {https://ui.adsabs.harvard.edu/abs/2021ApJ...917...40K} {917, 40}

\bibitem[\protect\citeauthoryear{{Lenoble}, {Commer{\c{c}}on}  \& {Rosdahl}}{{Lenoble} et~al.}{2024}]{Lenoble2024}
{Lenoble} R.,  {Commer{\c{c}}on} B.,   {Rosdahl} J.,  2024, \mn@doi [arXiv e-prints] {10.48550/arXiv.2401.16821}, \href {https://ui.adsabs.harvard.edu/abs/2024arXiv240116821L} {p. arXiv:2401.16821}

\bibitem[\protect\citeauthoryear{{Mardini}, {Frebel}  \& {Chiti}}{{Mardini} et~al.}{2024}]{Mardini2024}
{Mardini} M.~K.,  {Frebel} A.,   {Chiti} A.,  2024, \mn@doi [\mnras] {10.1093/mnrasl/slad197}, \href {https://ui.adsabs.harvard.edu/abs/2024MNRAS.529L..60M} {529, L60}

\bibitem[\protect\citeauthoryear{{Naoz} \& {Narayan}}{{Naoz} \& {Narayan}}{2014}]{Naoz2014}
{Naoz} S.,  {Narayan} R.,  2014, \mn@doi [\apjl] {10.1088/2041-8205/791/1/L8}, \href {https://ui.adsabs.harvard.edu/abs/2014ApJ...791L...8N} {791, L8}

\bibitem[\protect\citeauthoryear{{Nebrin}, {Giri}  \& {Mellema}}{{Nebrin} et~al.}{2023}]{Nebrin2023}
{Nebrin} O.,  {Giri} S.~K.,   {Mellema} G.,  2023, \mn@doi [\mnras] {10.1093/mnras/stad1852}, \href {https://ui.adsabs.harvard.edu/abs/2023MNRAS.524.2290N} {524, 2290}

\bibitem[\protect\citeauthoryear{{Nishijima}, {Hirano}  \& {Umeda}}{{Nishijima} et~al.}{2024}]{Nishijima2024}
{Nishijima} S.,  {Hirano} S.,   {Umeda} H.,  2024, \mn@doi [\apj] {10.3847/1538-4357/ad2fc9}, \href {https://ui.adsabs.harvard.edu/abs/2024ApJ...965..141N} {965, 141}

\bibitem[\protect\citeauthoryear{{Park}, {Ahn}, {Yoshida}  \& {Hirano}}{{Park} et~al.}{2020}]{Park2020btd}
{Park} H.,  {Ahn} K.,  {Yoshida} N.,   {Hirano} S.,  2020, \mn@doi [\apj] {10.3847/1538-4357/aba26e}, \href {https://ui.adsabs.harvard.edu/abs/2020ApJ...900...30P} {900, 30}

\bibitem[\protect\citeauthoryear{{Park}, {Shapiro}, {Ahn}, {Yoshida}  \& {Hirano}}{{Park} et~al.}{2021}]{Park2021btd}
{Park} H.,  {Shapiro} P.~R.,  {Ahn} K.,  {Yoshida} N.,   {Hirano} S.,  2021, \mn@doi [\apj] {10.3847/1538-4357/abd7f4}, \href {https://ui.adsabs.harvard.edu/abs/2021ApJ...908...96P} {908, 96}

\bibitem[\protect\citeauthoryear{{Placco} et~al.,}{{Placco} et~al.}{2016}]{Placco2016}
{Placco} V.~M.,  et~al., 2016, \mn@doi [\apj] {10.3847/0004-637X/833/1/21}, \href {https://ui.adsabs.harvard.edu/abs/2016ApJ...833...21P} {833, 21}

\bibitem[\protect\citeauthoryear{{Placco} et~al.,}{{Placco} et~al.}{2021}]{Placco2021}
{Placco} V.~M.,  et~al., 2021, \mn@doi [\apjl] {10.3847/2041-8213/abf93d}, \href {https://ui.adsabs.harvard.edu/abs/2021ApJ...912L..32P} {912, L32}

\bibitem[\protect\citeauthoryear{{Planck Collaboration} et~al.,}{{Planck Collaboration} et~al.}{2020}]{PLANCK2018}
{Planck Collaboration} et~al., 2020, \mn@doi [\aap] {10.1051/0004-6361/201833910}, \href {https://ui.adsabs.harvard.edu/abs/2020A&A...641A...6P} {641, A6}

\bibitem[\protect\citeauthoryear{{Robertson} et~al.,}{{Robertson} et~al.}{2024}]{Robertson2024}
{Robertson} B.,  et~al., 2024, \mn@doi [\apj] {10.3847/1538-4357/ad463d}, \href {https://ui.adsabs.harvard.edu/abs/2024ApJ...970...31R} {970, 31}

\bibitem[\protect\citeauthoryear{{Rossi}, {Salvadori}, {Sk{\'u}lad{\'o}ttir}, {Vanni}  \& {Koutsouridou}}{{Rossi} et~al.}{2024}]{Rossi2024}
{Rossi} M.,  {Salvadori} S.,  {Sk{\'u}lad{\'o}ttir} {\'A}.,  {Vanni} I.,   {Koutsouridou} I.,  2024, \mn@doi [arXiv e-prints] {10.48550/arXiv.2406.12960}, \href {https://ui.adsabs.harvard.edu/abs/2024arXiv240612960R} {p. arXiv:2406.12960}

\bibitem[\protect\citeauthoryear{{Schaerer}}{{Schaerer}}{2002}]{Schaerer2002}
{Schaerer} D.,  2002, \mn@doi [\aap] {10.1051/0004-6361:20011619}, \href {https://ui.adsabs.harvard.edu/abs/2002A&A...382...28S} {382, 28}

\bibitem[\protect\citeauthoryear{{Schauer}, {Glover}, {Klessen}  \& {Clark}}{{Schauer} et~al.}{2021a}]{Schauer2021SV+UV}
{Schauer} A. T.~P.,  {Glover} S. C.~O.,  {Klessen} R.~S.,   {Clark} P.,  2021a, \mn@doi [\mnras] {10.1093/mnras/stab1953}, \href {https://ui.adsabs.harvard.edu/abs/2021MNRAS.507.1775S} {507, 1775}

\bibitem[\protect\citeauthoryear{{Schauer}, {Bromm}, {Boylan-Kolchin}, {Glover}  \& {Klessen}}{{Schauer} et~al.}{2021b}]{Schauer2021SIGO}
{Schauer} A. T.~P.,  {Bromm} V.,  {Boylan-Kolchin} M.,  {Glover} S. C.~O.,   {Klessen} R.~S.,  2021b, \mn@doi [\apj] {10.3847/1538-4357/ac27aa}, \href {https://ui.adsabs.harvard.edu/abs/2021ApJ...922..193S} {922, 193}

\bibitem[\protect\citeauthoryear{{Schneider}, {Omukai}, {Limongi}, {Ferrara}, {Salvaterra}, {Chieffi}  \& {Bianchi}}{{Schneider} et~al.}{2012}]{Schneider2012}
{Schneider} R.,  {Omukai} K.,  {Limongi} M.,  {Ferrara} A.,  {Salvaterra} R.,  {Chieffi} A.,   {Bianchi} S.,  2012, \mn@doi [\mnras] {10.1111/j.1745-3933.2012.01257.x}, \href {https://ui.adsabs.harvard.edu/abs/2012MNRAS.423L..60S} {423, L60}

\bibitem[\protect\citeauthoryear{{Sk{\'u}lad{\'o}ttir} et~al.,}{{Sk{\'u}lad{\'o}ttir} et~al.}{2021}]{Skuladottir2021}
{Sk{\'u}lad{\'o}ttir} {\'A}.,  et~al., 2021, \mn@doi [\apjl] {10.3847/2041-8213/ac0dc2}, \href {https://ui.adsabs.harvard.edu/abs/2021ApJ...915L..30S} {915, L30}

\bibitem[\protect\citeauthoryear{{Springel}}{{Springel}}{2005}]{Springel2005}
{Springel} V.,  2005, \mn@doi [\mnras] {10.1111/j.1365-2966.2005.09655.x}, \href {https://ui.adsabs.harvard.edu/abs/2005MNRAS.364.1105S} {364, 1105}

\bibitem[\protect\citeauthoryear{{Stacy}, {Bromm}  \& {Loeb}}{{Stacy} et~al.}{2011}]{Stacy2011}
{Stacy} A.,  {Bromm} V.,   {Loeb} A.,  2011, \mn@doi [\apjl] {10.1088/2041-8205/730/1/L1}, \href {https://ui.adsabs.harvard.edu/abs/2011ApJ...730L...1S} {730, L1}

\bibitem[\protect\citeauthoryear{{Sugimura}, {Matsumoto}, {Hosokawa}, {Hirano}  \& {Omukai}}{{Sugimura} et~al.}{2023}]{Sugimura2023}
{Sugimura} K.,  {Matsumoto} T.,  {Hosokawa} T.,  {Hirano} S.,   {Omukai} K.,  2023, \mn@doi [\apj] {10.3847/1538-4357/ad02fc}, \href {https://ui.adsabs.harvard.edu/abs/2023ApJ...959...17S} {959, 17}

\bibitem[\protect\citeauthoryear{{Susa}}{{Susa}}{2019}]{Susa2019}
{Susa} H.,  2019, \mn@doi [\apj] {10.3847/1538-4357/ab1b6f}, \href {https://ui.adsabs.harvard.edu/abs/2019ApJ...877...99S} {877, 99}

\bibitem[\protect\citeauthoryear{{Tseliakhovich} \& {Hirata}}{{Tseliakhovich} \& {Hirata}}{2010}]{Tseliakhovich2010}
{Tseliakhovich} D.,  {Hirata} C.,  2010, \mn@doi [\prd] {10.1103/PhysRevD.82.083520}, \href {https://ui.adsabs.harvard.edu/abs/2010PhRvD..82h3520T} {82, 083520}

\bibitem[\protect\citeauthoryear{{Tseliakhovich}, {Barkana}  \& {Hirata}}{{Tseliakhovich} et~al.}{2011}]{Tseliakhovich2011}
{Tseliakhovich} D.,  {Barkana} R.,   {Hirata} C.~M.,  2011, \mn@doi [\mnras] {10.1111/j.1365-2966.2011.19541.x}, \href {https://ui.adsabs.harvard.edu/abs/2011MNRAS.418..906T} {418, 906}

\bibitem[\protect\citeauthoryear{{Uysal} \& {Hartwig}}{{Uysal} \& {Hartwig}}{2023}]{Uysal2023}
{Uysal} B.,  {Hartwig} T.,  2023, \mn@doi [\mnras] {10.1093/mnras/stad350}, \href {https://ui.adsabs.harvard.edu/abs/2023MNRAS.520.3229U} {520, 3229}

\bibitem[\protect\citeauthoryear{{Wise}, {Regan}, {O'Shea}, {Norman}, {Downes}  \& {Xu}}{{Wise} et~al.}{2019}]{Wise2019}
{Wise} J.~H.,  {Regan} J.~A.,  {O'Shea} B.~W.,  {Norman} M.~L.,  {Downes} T.~P.,   {Xu} H.,  2019, \mn@doi [\nat] {10.1038/s41586-019-0873-4}, \href {https://ui.adsabs.harvard.edu/abs/2019Natur.566...85W} {566, 85}

\bibitem[\protect\citeauthoryear{{Xing} et~al.,}{{Xing} et~al.}{2023}]{Xing2023}
{Xing} Q.-F.,  et~al., 2023, \mn@doi [\nat] {10.1038/s41586-023-06028-1}, \href {https://ui.adsabs.harvard.edu/abs/2023Natur.618..712X} {618, 712}

\bibitem[\protect\citeauthoryear{{Yoshida}, {Oh}, {Kitayama}  \& {Hernquist}}{{Yoshida} et~al.}{2007}]{Yoshida2007}
{Yoshida} N.,  {Oh} S.~P.,  {Kitayama} T.,   {Hernquist} L.,  2007, \mn@doi [\apj] {10.1086/518227}, \href {https://ui.adsabs.harvard.edu/abs/2007ApJ...663..687Y} {663, 687}

\bibitem[\protect\citeauthoryear{{Yoshida}, {Omukai}  \& {Hernquist}}{{Yoshida} et~al.}{2008}]{Yoshida2008}
{Yoshida} N.,  {Omukai} K.,   {Hernquist} L.,  2008, \mn@doi [Science] {10.1126/science.1160259}, \href {https://ui.adsabs.harvard.edu/abs/2008Sci...321..669Y} {321, 669}

\makeatother
\end{thebibliography}

%%%%%%%%%%%%%%%%%%%%%%%%%%%%%%%%%%%%%%%%%%%%%%%%%%
%%%%%%%%%%%%%%%%%%%% APPENDIX % %%%%%%%%%%%%%%%%%%
%%%%%%%%%%%%%%%%%%%%%%%%%%%%%%%%%%%%%%%%%%%%%%%%%%
\appendix

%%%%%%%%%%%%%%%%%%%%%%%%%%%%%%%%%%%%%%%%%%%%%%%%%%
\section{Results of 120 models}
\label{app:table_results_each}
%%%%%%%%%%%%%%%%%%%%%%%%%%%%%%%%%%%%%%%%%%%%%%%%%%

%%%%%%%%%%%%%%%%%%%%%%%%%%%%%%%%%%%%%%%%%%%%%%%%%%
\begin{table*}
\centering
\caption{Parameters and results of 120 models}
\label{table:each}
\begin{tabular}{@{}lccrccccrrcrrr@{}}
\hline
Model & $\vsvsigma$ & $z$ & $\Rvir$ & $\Mvir$ & $\fbaryon$ & Class & HD & $\Ncore$ & $\Mcoretot$ & $\epsIII$ & $\Mcorefirst$ & $\Mcoresecond$ & $\qcore$ \\
& & & (pc) & ($\msun$) & & & & & ($\msun$) & & ($\msun$) & ($\msun$) & \\
\hline
I01V00 & 0.0 & 36.80 &  63.1 & $3.723\times10^5$ & 0.109 & {\it High} &   &  1 &   3882 & 0.0958 &  3882 &     - &     - \\
I01V10 & 1.0 & 30.91 & 125.9 & $1.920\times10^6$ & 0.106 & & Y &  1 &   9567 & 0.0469 &  9567 &     - &     - \\
I01V15 & 1.5 & 27.45 & 223.9 & $7.856\times10^6$ & 0.114 & & Y &  2 &   5253 & 0.0059 &  5122 &   131 & 0.026 \\
I01V20 & 2.0 & 25.84 & 316.2 & $1.794\times10^7$ & 0.118 & &   &  5 &  45957 & 0.0218 & 29131 &  6038 & 0.207 \\
I01V25 & 2.5 & 25.10 & 354.8 & $2.419\times10^7$ & 0.109 & &   &  1 &  46198 & 0.0175 & 46198 &     - &     - \\
I01V30 & 3.0 & 23.33 & 562.3 & $6.468\times10^7$ & 0.121 & &   &  7 &  42438 & 0.0054 & 33401 &  3292 & 0.099 \\
\hline
I02V00 & 0.0 & 34.09 & 100.0 & $1.319\times10^6$ & 0.121 & {\it High} &   &  1 &   8635 & 0.0542 &  8635 &     - &     - \\
I02V10 & 1.0 & 31.53 & 141.3 & $2.855\times10^6$ & 0.108 & &   &  1 &  19714 & 0.0638 & 19714 &     - &     - \\
I02V15 & 1.5 & 30.60 & 158.5 & $3.576\times10^6$ & 0.100 & &   &  7 &  18935 & 0.0528 & 14945 &  1765 & 0.118 \\
I02V20 & 2.0 & 29.84 & 177.8 & $4.163\times10^6$ & 0.096 & &   &  2 &  19186 & 0.0481 & 16122 &  3065 & 0.190 \\
I02V25 & 2.5 & 21.28 & 501.2 & $3.670\times10^7$ & 0.146 & &   &  4 &  60304 & 0.0113 & 22010 & 20668 & 0.939 \\
I02V30 & 3.0 & 19.50 & 562.3 & $4.732\times10^7$ & 0.169 & &   &  2 &  33990 & 0.0043 & 23552 & 10439 & 0.443 \\
\hline
I03V00 & 0.0 & 31.68 &  63.1 & $2.195\times10^5$ & 0.111 & {\it Middle} &   &  1 &   3158 & 0.1295 &  3158 &     - &     - \\
I03V10 & 1.0 & 20.03 & 354.8 & $1.169\times10^7$ & 0.125 & &   &  3 &  47059 & 0.0321 & 41683 &  5276 & 0.127 \\
I03V15 & 1.5 & 21.37 & 316.2 & $9.943\times10^6$ & 0.127 & {\it E1} & Y &  1 &   7381 & 0.0058 &  7381 &     - &     - \\
I03V20 & 2.0 & 20.81 & 354.8 & $1.229\times10^7$ & 0.124 & &   & 15 &  69239 & 0.0455 & 44308 &  6124 & 0.138 \\
I03V25 & 2.5 & 20.41 & 354.8 & $1.341\times10^7$ & 0.125 & &   &  6 &  49196 & 0.0293 & 26382 & 21452 & 0.813 \\
I03V30 & 3.0 & 20.13 & 398.1 & $1.531\times10^7$ & 0.124 & &   &  3 &  26778 & 0.0141 & 25882 &   493 & 0.019 \\
\hline
I04V00 & 0.0 & 30.53 & 100.0 & $7.717\times10^5$ & 0.116 & {\it Middle} &   &  2 &   4560 & 0.0511 &  4441 &   119 & 0.027 \\
I04V10 & 1.0 & 30.22 & 100.0 & $7.972\times10^5$ & 0.114 & &   &  1 &   2781 & 0.0306 &  2781 &     - &     - \\
I04V15 & 1.5 & 27.45 & 141.3 & $1.647\times10^6$ & 0.128 & & Y &  8 &   5912 & 0.0280 &  3203 &   997 & 0.311 \\
I04V20 & 2.0 & 24.82 & 251.2 & $6.800\times10^6$ & 0.109 & &   &  1 &   2243 & 0.0030 &  2243 &     - &     - \\
I04V25 & 2.5 & 23.02 & 316.2 & $1.183\times10^7$ & 0.113 & &   &  2 &  13813 & 0.0103 &  9544 &  4269 & 0.447 \\
I04V30 & 3.0 & 23.31 & 281.8 & $8.841\times10^6$ & 0.089 & {\it E1} &   &  1 &  21215 & 0.0270 & 21215 &     - &     - \\
\hline
I05V00 & 0.0 & 29.63 &  89.1 & $6.011\times10^5$ & 0.123 & {\it Middle} &   &  1 &   3081 & 0.0416 &  3081 &     - &     - \\
I05V10 & 1.0 & 26.96 & 177.8 & $3.807\times10^6$ & 0.116 & & Y &  1 &   4500 & 0.0102 &  4500 &     - &     - \\
I05V15 & 1.5 & 25.45 & 223.9 & $6.182\times10^6$ & 0.112 & &   &  9 &  56412 & 0.0818 & 23528 & 16669 & 0.708 \\
I05V20 & 2.0 & 24.45 & 281.8 & $9.226\times10^6$ & 0.106 & &   &  8 &  71629 & 0.0734 & 46693 & 18986 & 0.407 \\
I05V25 & 2.5 & 23.71 & 316.2 & $1.288\times10^7$ & 0.100 & &   &  7 &  45237 & 0.0351 & 33732 &  5872 & 0.174 \\
I05V30 & 3.0 & 23.05 & 354.8 & $1.571\times10^7$ & 0.100 & &   &  3 &  17258 & 0.0110 &  6948 &  6002 & 0.864 \\
\hline
I06V00 & 0.0 & 29.03 &  89.1 & $4.698\times10^5$ & 0.119 & {\it Middle} &   &  1 &   2404 & 0.0429 &  2404 &     - &     - \\
I06V10 & 1.0 & 25.01 & 125.9 & $1.079\times10^6$ & 0.119 & & Y &  1 &   3981 & 0.0309 &  3981 &     - &     - \\
I06V15 & 1.5 & 23.97 & 199.5 & $3.465\times10^6$ & 0.111 & & Y &  2 &   8391 & 0.0219 &  6766 &  1624 & 0.240 \\
I06V20 & 2.0 & 21.73 & 316.2 & $9.533\times10^6$ & 0.122 & &   &  3 &  15988 & 0.0137 & 11579 &  3003 & 0.259 \\
I06V25 & 2.5 & 20.62 & 354.8 & $1.222\times10^7$ & 0.116 & &   &  3 &   9198 & 0.0065 &  7232 &  1568 & 0.217 \\
I06V30 & 3.0 & 19.63 & 631.0 & $6.869\times10^7$ & 0.117 & &   & 11 &  84097 & 0.0105 & 44502 & 10855 & 0.244 \\
\hline
I07V00 & 0.0 & 28.64 & 112.2 & $1.106\times10^6$ & 0.129 & {\it Middle} & Y &  2 &   3000 & 0.0210 &  2050 &   949 & 0.463 \\
I07V10 & 1.0 & 26.79 & 158.5 & $2.514\times10^6$ & 0.117 & &   &  2 &  11161 & 0.0380 &  6444 &  4717 & 0.732 \\
I07V15 & 1.5 & 21.61 & 398.1 & $2.030\times10^7$ & 0.110 & &   &  6 &  45839 & 0.0206 & 18220 & 17657 & 0.969 \\
I07V20 & 2.0 & 23.21 & 354.8 & $1.731\times10^7$ & 0.114 & {\it E1} &   &  7 &  57912 & 0.0294 & 31496 &  9606 & 0.305 \\
I07V25 & 2.5 & 22.55 & 354.8 & $1.822\times10^7$ & 0.109 & &   & 20 &  63743 & 0.0320 & 32817 & 12111 & 0.369 \\
I07V30 & 3.0 & 22.57 & 354.8 & $1.705\times10^7$ & 0.109 & {\it E1} & Y &  1 &   7494 & 0.0040 &  7494 &     - &     - \\
\hline
\end{tabular}
\begin{flushleft}
{\it Notes.} 
Same as Table~\ref{table:average} but for each model.
Column 1: model name.
Models I01, I04, I10, I11, I13, I14, and I19 correspond to Halos A, B, C, D, E, F, and G in \citetalias{Hirano2023PaperI}.
Column 4: radius ($\Rvir$) at the virial scale.
We determine the virial radii from distance grids, divided into 20 grids with one order of magnitude of distance on a logarithmic scale, so the same values appear in the table.
Column 7: Class names as three classes for each model series ({\it High}, {\it Middle}, and {\it Low} as described in Section~\ref{sec:res}) in columns of no-SV models whereas three exceptional SV dependence on $z$-$\Mvir$ diagram ({\it E1}, {\it E2}, and {\it E3} as described in Section~\ref{sec:res:virial:exception}) in columns of models with SV.
Column 8: whether HD-cooling is enabled (abundance ratio criterion, $f_{\rm HD}/f_{\rm H_2} \geq 10^{-3}$) at the end of the calculation.
Column 13: mass of the secondary core ($\Mcoresecond$).
Column 14: mass ratio of the primary and secondary core ($\qcore=\Mcoresecond/\Mcorefirst$).
There is no data in columns 13 and 14 for models with $\Ncore = 1$ because there is no secondary core.
\end{flushleft}
\end{table*}

\begin{table*}
\centering
\contcaption{
%A table continued from the previous one.
}
\begin{tabular}{@{}lccrccccrrcrrr@{}}
\hline
Model & $\vsvsigma$ & $z$ & $\Rvir$ & $\Mvir$ & $\fbaryon$ & Class & HD & $\Ncore$ & $\Mcoretot$ & $\epsIII$ & $\Mcorefirst$ & $\Mcoresecond$ & $\qcore$ \\
& & & (pc) & ($\msun$) & & & & & ($\msun$) & & ($\msun$) & ($\msun$) & \\
\hline
I08V00 & 0.0 & 28.42 & 158.5 & $2.637\times10^6$ & 0.130 & {\it High} &   &  1 &  30352 & 0.0887 & 30352 &     - &     - \\
I08V10 & 1.0 & 27.91 & 158.5 & $2.786\times10^6$ & 0.132 & &   &  1 &  14592 & 0.0397 & 14592 &     - &     - \\
I08V15 & 1.5 & 25.85 & 251.2 & $8.700\times10^6$ & 0.128 & &   &  5 &  57294 & 0.0515 & 52925 &  1516 & 0.029 \\
I08V20 & 2.0 & 25.01 & 316.2 & $1.543\times10^7$ & 0.115 & &   & 23 &  83535 & 0.0469 & 26394 & 15624 & 0.592 \\
I08V25 & 2.5 & 24.31 & 354.8 & $1.980\times10^7$ & 0.106 & &   &  1 &  32946 & 0.0157 & 32946 &     - &     - \\
I08V30 & 3.0 & 23.55 & 398.1 & $2.587\times10^7$ & 0.097 & &   &  3 &  12589 & 0.0050 &  6296 &  5708 & 0.907 \\
\hline
I09V00 & 0.0 & 28.03 & 158.5 & $2.503\times10^6$ & 0.133 & {\it High} &   &  1 &  24920 & 0.0750 & 24920 &     - &     - \\
I09V10 & 1.0 & 29.63 & 112.2 & $1.061\times10^6$ & 0.118 & {\it E1} &   &  1 &   5025 & 0.0402 &  5025 &     - &     - \\
I09V15 & 1.5 & 26.45 & 199.5 & $4.638\times10^6$ & 0.121 & &   &  5 &  13707 & 0.0245 &  6391 &  5219 & 0.817 \\
I09V20 & 2.0 & 25.16 & 251.2 & $7.538\times10^6$ & 0.118 & &   &  5 &  34013 & 0.0381 & 21769 &  5673 & 0.261 \\
I09V25 & 2.5 & 24.21 & 281.8 & $1.002\times10^7$ & 0.118 & &   &  2 &   4491 & 0.0038 &  2854 &  1637 & 0.574 \\
I09V30 & 3.0 & 23.48 & 316.2 & $1.304\times10^7$ & 0.128 & &   &  1 &  31843 & 0.0191 & 31843 &     - &     - \\
\hline
I10V00 & 0.0 & 27.61 & 112.2 & $9.544\times10^5$ & 0.143 & {\it Middle} & Y &  4 &   3309 & 0.0242 &  2824 &   269 & 0.095 \\
I10V10 & 1.0 & 25.38 & 177.8 & $2.608\times10^6$ & 0.138 & &   &  1 &   9258 & 0.0258 &  9258 &     - &     - \\
I10V15 & 1.5 & 24.20 & 199.5 & $3.801\times10^6$ & 0.130 & &   &  1 &   4787 & 0.0097 &  4787 &     - &     - \\
I10V20 & 2.0 & 23.04 & 251.2 & $6.784\times10^6$ & 0.132 & & Y &  2 &  12005 & 0.0134 & 11571 &   434 & 0.037 \\
I10V25 & 2.5 & 21.37 & 398.1 & $2.009\times10^7$ & 0.148 & &   &  9 &  42078 & 0.0141 & 16271 & 13344 & 0.820 \\
I10V30 & 3.0 & 20.79 & 446.7 & $2.721\times10^7$ & 0.134 & &   & 14 &  92182 & 0.0252 & 59830 &  6930 & 0.116 \\
\hline
I11V00 & 0.0 & 27.60 & 158.5 & $2.675\times10^6$ & 0.127 & {\it High} &   &  1 &   7619 & 0.0223 &  7619 &     - &     - \\
I11V10 & 1.0 & 25.42 & 223.9 & $5.918\times10^6$ & 0.140 & &   &  2 &  11876 & 0.0143 &  7787 &  4089 & 0.525 \\
I11V15 & 1.5 & 24.51 & 281.8 & $1.113\times10^7$ & 0.105 & &   &  4 &  40245 & 0.0346 & 23696 & 12177 & 0.514 \\
I11V20 & 2.0 & 23.75 & 316.2 & $1.343\times10^7$ & 0.116 & &   &  3 &  11358 & 0.0073 &  7660 &  2459 & 0.321 \\
I11V25 & 2.5 & 21.09 & 501.2 & $4.029\times10^7$ & 0.115 & &   & 20 &  99898 & 0.0216 & 19532 &  9357 & 0.479 \\
I11V30 & 3.0 & 20.85 & 501.2 & $4.055\times10^7$ & 0.108 & &   & 25 &  71040 & 0.0162 & 21052 & 16954 & 0.805 \\
\hline
I12V00 & 0.0 & 26.61 & 100.0 & $6.222\times10^5$ & 0.142 & {\it Low} & Y &  4 &   4474 & 0.0506 &  3249 &   769 & 0.237 \\
I12V10 & 1.0 & 22.58 & 177.8 & $2.168\times10^6$ & 0.132 & & Y &  1 &   1172 & 0.0041 &  1172 &     - &     - \\
I12V15 & 1.5 & 20.31 & 281.8 & $6.683\times10^6$ & 0.129 & &   &  1 &   5466 & 0.0064 &  5466 &     - &     - \\
I12V20 & 2.0 & 19.09 & 354.8 & $1.027\times10^7$ & 0.117 & &   &  5 & 102428 & 0.0856 & 57251 & 40004 & 0.699 \\
I12V25 & 2.5 & 18.67 & 354.8 & $1.052\times10^7$ & 0.105 & &   &  3 &  22486 & 0.0203 & 20588 &  1491 & 0.072 \\
I12V30 & 3.0 & 18.13 & 398.1 & $1.242\times10^7$ & 0.101 & &   &  4 &  15564 & 0.0125 &  8525 &  3612 & 0.424 \\
\hline
I13V00 & 0.0 & 26.43 & 112.2 & $7.845\times10^5$ & 0.141 & {\it Low} &   &  1 &   8464 & 0.0765 &  8464 &     - &     - \\
I13V10 & 1.0 & 23.46 & 177.8 & $2.268\times10^6$ & 0.144 & &   &  3 &  19317 & 0.0590 & 10009 &  5865 & 0.586 \\
I13V15 & 1.5 & 21.70 & 251.2 & $4.545\times10^6$ & 0.118 & &   &  2 &  18468 & 0.0345 & 11888 &  6579 & 0.553 \\
I13V20 & 2.0 & 19.55 & 316.2 & $7.821\times10^6$ & 0.112 & &   &  1 &   8363 & 0.0095 &  8363 &     - &     - \\
I13V25 & 2.5 & 17.93 & 398.1 & $1.259\times10^7$ & 0.119 & &   &  2 &  18841 & 0.0125 & 16733 &  2151 & 0.129 \\
I13V30 & 3.0 & 17.23 & 446.7 & $1.652\times10^7$ & 0.114 & &   &  1 &  13916 & 0.0074 & 13916 &     - &     - \\
\hline
I14V00 & 0.0 & 25.42 & 141.3 & $1.471\times10^6$ & 0.148 & {\it Middle} &   &  2 &   9274 & 0.0427 &  6777 &  2497 & 0.368 \\
I14V10 & 1.0 & 22.70 & 223.9 & $4.561\times10^6$ & 0.138 & &   &  5 &  55739 & 0.0886 & 23849 & 18663 & 0.783 \\
I14V15 & 1.5 & 22.21 & 251.2 & $5.590\times10^6$ & 0.127 & &   &  4 &  25121 & 0.0354 & 24568 &   175 & 0.007 \\
I14V20 & 2.0 & 21.27 & 281.8 & $7.451\times10^6$ & 0.120 & &   &  1 &  22094 & 0.0248 & 22094 &     - &     - \\
I14V25 & 2.5 & 20.05 & 354.8 & $1.134\times10^7$ & 0.120 & &   &  1 &   7717 & 0.0057 &  7717 &     - &     - \\
I14V30 & 3.0 & 19.79 & 398.1 & $1.468\times10^7$ & 0.124 & &   &  2 &  21315 & 0.0117 & 13043 &  8271 & 0.634 \\
\hline
\end{tabular}
\end{table*}

\begin{table*}
\centering
\contcaption{
%A table continued from the previous one.
}
\begin{tabular}{@{}lccrccccrrcrrr@{}}
\hline
Model & $\vsvsigma$ & $z$ & $\Rvir$ & $\Mvir$ & $\fbaryon$ & Class & HD & $\Ncore$ & $\Mcoretot$ & $\epsIII$ & $\Mcorefirst$ & $\Mcoresecond$ & $\qcore$ \\
& & & (pc) & ($\msun$) & & & & & ($\msun$) & & ($\msun$) & ($\msun$) & \\
\hline
I15V00 & 0.0 & 24.71 & 125.9 & $1.016\times10^6$ & 0.135 & {\it Middle} &   &  2 &   9822 & 0.0717 &  8069 &  1753 & 0.217 \\
I15V10 & 1.0 & 24.67 & 199.5 & $3.372\times10^6$ & 0.121 & &   &  2 &  12668 & 0.0311 &  8947 &  3721 & 0.416 \\
I15V15 & 1.5 & 19.74 & 446.7 & $2.000\times10^7$ & 0.131 & &   &  3 &  31820 & 0.0121 & 19821 &  7676 & 0.387 \\
I15V20 & 2.0 & 20.58 & 398.1 & $1.615\times10^7$ & 0.130 & {\it E1} &   &  1 &  10809 & 0.0052 & 10809 &     - &     - \\
I15V25 & 2.5 & 20.22 & 446.7 & $2.315\times10^7$ & 0.123 & &   &  1 &  19353 & 0.0068 & 19353 &     - &     - \\
I15V30 & 3.0 & 20.39 & 501.2 & $3.070\times10^7$ & 0.115 & {\it E2} &   &  5 &  58686 & 0.0166 & 40081 &  6353 & 0.158 \\
\hline
I16V00 & 0.0 & 23.82 & 112.2 & $6.588\times10^5$ & 0.143 & {\it Low} &   &  1 &   5635 & 0.0599 &  5635 &     - &     - \\
I16V10 & 1.0 & 21.55 & 177.8 & $1.797\times10^6$ & 0.139 & &   &  2 &  19840 & 0.0795 & 19392 &   448 & 0.023 \\
I16V15 & 1.5 & 18.98 & 316.2 & $7.158\times10^6$ & 0.129 & &   &  1 &  29981 & 0.0324 & 29981 &     - &     - \\
I16V20 & 2.0 & 17.43 & 501.2 & $2.024\times10^7$ & 0.143 & &   &  4 &  24095 & 0.0083 & 15335 &  7307 & 0.477 \\
I16V25 & 2.5 & 16.96 & 562.3 & $2.577\times10^7$ & 0.131 & &   &  3 &  30955 & 0.0091 & 25432 &  5394 & 0.212 \\
I16V30 & 3.0 & 16.73 & 562.3 & $2.520\times10^7$ & 0.123 & {\it E3} &   &  1 &  28997 & 0.0093 & 28997 &     - &     - \\
\hline
I17V00 & 0.0 & 22.79 & 100.0 & $3.841\times10^5$ & 0.138 & {\it Low} &   &  1 &   5141 & 0.0972 &  5141 &     - &     - \\
I17V10 & 1.0 & 18.06 & 316.2 & $5.876\times10^6$ & 0.144 & &   &  1 &   7044 & 0.0083 &  7044 &     - &     - \\
I17V15 & 1.5 & 17.50 & 316.2 & $6.021\times10^6$ & 0.132 & &   &  1 &   4376 & 0.0055 &  4376 &     - &     - \\
I17V20 & 2.0 & 17.72 & 281.8 & $4.470\times10^6$ & 0.143 & {\it E1} &   &  1 &   4303 & 0.0067 &  4303 &     - &     - \\
I17V25 & 2.5 & 17.25 & 316.2 & $5.552\times10^6$ & 0.130 & &   &  3 &  21318 & 0.0296 & 20138 &  1108 & 0.055 \\
I17V30 & 3.0 & 16.96 & 316.2 & $5.597\times10^6$ & 0.137 & &   &  3 &  13186 & 0.0173 & 12930 &   177 & 0.014 \\
\hline
I18V00 & 0.0 & 22.02 & 223.9 & $3.695\times10^6$ & 0.150 & {\it Low} &   &  1 &   4390 & 0.0079 &  4390 &     - &     - \\
I18V10 & 1.0 & 22.30 & 199.5 & $2.768\times10^6$ & 0.148 & {\it E1} & Y &  1 &   6007 & 0.0146 &  6007 &     - &     - \\
I18V15 & 1.5 & 21.27 & 223.9 & $3.585\times10^6$ & 0.143 & & Y &  1 &   5512 & 0.0107 &  5512 &     - &     - \\
I18V20 & 2.0 & 19.33 & 354.8 & $1.058\times10^7$ & 0.139 & & Y & 11 &  11138 & 0.0076 &  6821 &  1151 & 0.169 \\
I18V25 & 2.5 & 18.23 & 446.7 & $1.702\times10^7$ & 0.135 & &   &  3 &  32918 & 0.0143 & 29685 &  3111 & 0.105 \\
I18V30 & 3.0 & 18.02 & 446.7 & $1.768\times10^7$ & 0.130 & &   &  3 &   7632 & 0.0033 &  3468 &  2118 & 0.611 \\
\hline
I19V00 & 0.0 & 21.08 & 177.8 & $1.602\times10^6$ & 0.147 & {\it Low} & Y &  1 &   8488 & 0.0361 &  8488 &     - &     - \\
I19V10 & 1.0 & 20.98 & 177.8 & $1.563\times10^6$ & 0.146 & {\it E3}  & Y &  1 &   2814 & 0.0124 &  2814 &     - &     - \\
I19V15 & 1.5 & 19.19 & 223.9 & $2.700\times10^6$ & 0.159 & &   &  1 &   4675 & 0.0109 &  4675 &     - &     - \\
I19V20 & 2.0 & 17.33 & 398.1 & $9.673\times10^6$ & 0.146 & &   &  3 &   4608 & 0.0033 &  2599 &  1066 & 0.410 \\
I19V25 & 2.5 & 17.01 & 398.1 & $1.038\times10^7$ & 0.137 & &   &  1 &   6739 & 0.0047 &  6739 &     - &     - \\
I19V30 & 3.0 & 16.00 & 446.7 & $1.360\times10^7$ & 0.137 & &   &  2 &   2936 & 0.0016 &  1914 &  1022 & 0.534 \\
\hline
I20V00 & 0.0 & 16.52 & 398.1 & $8.400\times10^6$ & 0.151 & {\it Low} &   &  2 &  20101 & 0.0158 & 11922 &  8179 & 0.686 \\
I20V10 & 1.0 & 17.74 & 251.2 & $3.178\times10^6$ & 0.128 & {\it E1} & Y &  1 &   4394 & 0.0108 &  4394 &     - &     - \\
I20V15 & 1.5 & 16.86 & 316.2 & $4.375\times10^6$ & 0.160 & &   &  1 &  16003 & 0.0228 & 16003 &     - &     - \\
I20V20 & 2.0 & 15.61 & 354.8 & $5.969\times10^6$ & 0.151 & &   &  2 &  25799 & 0.0286 & 24654 &  1146 & 0.046 \\
I20V25 & 2.5 & 15.79 & 354.8 & $6.027\times10^6$ & 0.128 & {\it E2} &   &  1 &   5324 & 0.0069 &  5324 &     - &     - \\
I20V30 & 3.0 & 16.49 & 316.2 & $4.599\times10^6$ & 0.134 & {\it E1} &   &  1 &   3205 & 0.0052 &  3205 &     - &     - \\
\hline
\end{tabular}
\end{table*}
%%%%%%%%%%%%%%%%%%%%%%%%%%%%%%%%%%%%%%%%%%%%%%%%%%

\end{CJK}%! To show the Japanese language.

%%%%%%%%%%%%%%%%%%%%%%%%%%%%%%%%%%%%%%%%%%%%%%%%%%
% Don't change these lines

\bsp	% typesetting comment
\label{lastpage}
\end{document}